\documentclass[a4paper,prx,twocolumn,superscriptaddress,amsmath,amssymb,mathrsfs]{revtex4-2}
\usepackage{tabularx}
\usepackage{graphicx}
\usepackage{braket}
\usepackage{dcolumn}
\usepackage{bm}
\usepackage{epsfig}
\usepackage{rotating}
\usepackage{color}
\usepackage{hyperref}
\hypersetup{colorlinks=true,linkcolor=blue,citecolor=blue,urlcolor=blue}

\begin{document}

\newcommand{\minusone}{{-\!1}}

\title{Complete Controllability Despite Degeneracy:\\
Quantum Control of Enantiomer-Specific State Transfer in Chiral Molecules}

\author{M. Leibscher}
\thanks{These authors have contributed equally.}
\affiliation{Theoretische Physik, Universit\"{a}t Kassel,
Heinrich-Plett-Stra{\ss}e 40, 34132 Kassel, Germany}

\author{E. Pozzoli}
\thanks{These authors have contributed equally.}
\affiliation{
Laboratoire Jacques-Louis Lions, Sorbonne Universit\'e, Universit\'e de Paris, CNRS, Inria, Paris, France}

\author{C. P\'erez}
\affiliation{
  Deutsches Elektronen-Synchrotron (DESY), Notkestraße 85, 22607 Hamburg, Germany 
}

\author{M. Schnell}
\affiliation{
  Deutsches Elektronen-Synchrotron (DESY), Notkestraße 85, 22607 Hamburg, Germany 
}
\affiliation{
Institute of Physical Chemistry, Christian-Albrechts-Universit\"at zu Kiel, Max-Eyth-Straße 1, 24118 Kiel, Germany
}

\author{M. Sigalotti} 
\affiliation{
Laboratoire Jacques-Louis Lions, Sorbonne Universit\'e, Universit\'e de Paris, CNRS, Inria, Paris, France}

\author{U. Boscain} 
\affiliation{
Laboratoire Jacques-Louis Lions, Sorbonne Universit\'e, Universit\'e de Paris, CNRS, Inria, Paris, France}

\author{C. P. Koch} 
\email{christiane.koch@fu-berlin.de}
\affiliation{Theoretische Physik, Universit\"{a}t Kassel,
Heinrich-Plett-Stra{\ss}e 40, 34132 Kassel, Germany}
\affiliation{Dahlem Center for Complex Quantum Systems and Fachbereich Physik, Freie Universit\"{a}t Berlin,
Arnimallee 14, 14195 Berlin, Germany}

\date{\today}
 
\begin{abstract}
  We prove complete controllability for rotational states of an asymmetric top molecule belonging to degenerate values of the orientational quantum number $M$. Based on this insight, we construct a pulse sequence that energetically separates population initially distributed over degenerate $M$-states, as a precursor for orientational purification. 
  Introducing the concept of enantio-selective controllability, we determine the conditions for complete enantiomer-specific population transfer in chiral molecules and construct pulse sequences realizing this transfer for population initially distributed over degenerate $M$-states. This degeneracy presently limits enantiomer-selectivity for any initial state except the rotational ground state. Our work thus shows how to overcome an important obstacle towards separating, with electric fields only, left-handed from right-handed molecules in a racemic mixture.
\end{abstract}

\maketitle

\section{Introduction}

Molecular chirality ---
the fact that a chiral molecule cannot be superimposed with its mirror image by rotations and translations ---
is as ubiquitous as it is intriguing. 
The left-handed and right-handed versions  of a chiral molecule share almost all of their physical properties. Yet, the chemical and biological behavior of the two enantiomers typically differs dramatically. Detection of chirality and the ability to separate enantiomers therefore play a central role across the natural sciences.
To this end, chiroptical spectroscopy, the interrogation of chiral molecules with electromagnetic radiation~\cite{ChiroptBook}, has been a method of choice since the very discovery of molecular chirality.
For the detection of enantiomeric excess, for example, 
several new techniques  have recently been brought forward, including resonant phase-sensitive microwave three-wave mixing~\cite{PattersonNature13,ShubertAngewandte14,LobsigerJPCL15} and ultrafast spectroscopies based on photoelectron circular dichroism~\cite{FanoodNatureComm15,KastnerChemPhysChem16,CombyNatureComm18} or high-harmonic generation~\cite{BakyushevaPRX18,NeufeldPRX19}. They share, as a common feature, a sufficiently high sensitivity allowing for application in gas phase samples of randomly oriented molecules.

In contrast to the advances in chiroptical spectroscopy, the separation of enantiomers in a racemic mixture with electromagnetic fields is still an open challenge. A precursor to enantiomer separation is enantiomer-selective population excitation that transfers enantiomers in a mixture to two  different energy levels~\cite{EibenbergerPRL17,PerezAngewandte17,PerezJPCL18}. If the efficiency of the population transfer can be brought close to 100\%, an enantiopure sample can be distilled out of the racemate by e.g. ionizing all molecules in one of the two levels.
High-resolution searches of parity violation would also benefit from enantiomer-selective population excitation~\cite{EibenbergerPRL17,QuackARPC08,Hutzler20}.
However, at most a few per cent of enantiomer selectivity have so far been observed~\cite{EibenbergerPRL17,PerezAngewandte17,PerezJPCL18} in experiments that were all based on resonant phase-sensitive microwave three-wave mixing~\cite{PattersonNature13,ShubertAngewandte14,LobsigerJPCL15}. 
This is one of several chiroptical techniques based exclusively on light-matter interactions in the electric dipole approximation. An enantiomer-selective observable arises as a triple product of molecule-specific vectors which changes sign under exchange of the two enantiomers, independent of the molecular orientation~\cite{BychkovJETP01,OrdonezPRA18}.
In case of resonant microwave three-wave mixing, the triple product is formed by the Cartesian projections of the molecule's electric dipole moment onto the molecular frame~\cite{OrdonezPRA18,HirotaPJA12,GrabowAngewandte2013}.
Exploiting this fact in cyclic population transfer, enantiomer-selective population excitation is achieved by creating destructive interference for molecules of one handedness and constructive interference for the other handedness~\cite{KralPRL01}.

When using cyclic population transfer for enantiomer-selective population excitation~\cite{EibenbergerPRL17,PerezAngewandte17,PerezJPCL18}, two factors have been limiting the efficiency. One is the temperature of the sample or, more precisely, thermal population in the excited states targeted by the three-wave mixing. A solution to this problem simply consists in addressing levels which are sufficiently highly excited such that their thermal population vanishes~\cite{Leibscher19,Zhang20}. The second limitation is due to degeneracies within the rotational spectrum. Denoting the rotational quantum number by $J$, every energy level of a rigid asymmetric top rotor consists of $2J+1$ states with different values of the orientational quantum number $M$. While theoretical descriptions of resonant microwave three-wave mixing most often ignore the presence of degenerate energy levels~\cite{ShapiroPRL00,KralPRL01,LiPRL07,LiJCP10,PattersonNature13,HirotaPJA12,Zhang20,Vitanov19}, 
they yield correct predictions on transfer efficiency only for cycles which start from the non-degenerate rotational ground state ($J=0$)~\cite{LehmannJCP18,Leibscher19,Ye20}. Otherwise, cyclic excitation between three rotational energy levels involves a number of coupled, partially incomplete three-level systems~\cite{EibenbergerPRL17,PerezAngewandte17,PerezJPCL18}. This limits efficiency of enantiomer-selective population transfer, even in the absence of thermal population in the excited states.

Here, we solve this problem
in terms of both the basic light-matter couplings that are required as well as specific pulse sequences that achieve complete enantiomer-selective population transfer in degenerate rotational levels. In passing, we also obtain pulse sequences that energetically separate population that initially is incoherently distributed over degenerate rotational levels. This can be used, for example, as a precursor for distilling a specific orientation.
Key to our solution is a rigorous analysis of the controllability of asymmetric tops which builds on recent advances in the controllability of quantum rotors~\cite{BCCS,BCS,CS,Boscain19}.
Introducing the concept of enantiomer-sensitive controllability, 
we prove that complete enantiomer-specific population transfer is possible despite the $M$-degeneracy. To realize complete population transfer, the three-wave mixing pulse sequence has to be amended to consist of at least five different combinations of the three polarization directions and three frequencies. The corresponding modified three-wave mixing cycles are closed for all levels in the degenerate manifold, avoiding population loss, and they can easily be synchronized for complete population transfer, despite $M$-dependent Rabi frequencies.
Our work shows how to exploit recent mathematical insight~\cite{BCS,Boscain19} challenging the traditionally held belief that degeneracy of quantum states is associated with lack of controllability~\cite{ShapiroBook} in a practical physical application. In the example of microwave three-wave mixing spectroscopy, it is the 3D nature of the molecular geometry that is at the core of the degeneracy, and we find that complete controllability can be engineered by fully taking advantage of the 3D nature of the light-matter coupling.

The paper is organized as follows. In Sec.~\ref{sec:model}, we recall the properties of rigid asymmetric top molecules
and define the control problem arising in microwave three-wave mixing due to the orientational degeneracy. We briefly summarize the methods for controllability analysis in Sec.~\ref{sec:Lie},
applying them to rotational subsystems of an asymmetric top in Sec.~\ref{sec:controllabilityresults}.  While Sec.~\ref{sec:Lie} and ~\ref{sec:controllabilityresults} show how to leverage mathematical controllability analysis for the solution of practical control problems, readers interested only in the modified three-wave mixing pulse sequences may skip directly to Sec.~\ref{sec:carvone}, where we present pulse sequences yielding complete  enantiomer-specific population transfer for the example of carvone molecules \cite{PattersonPCCP14}.
We formulate general design principles for pulses driving orientation-selective, respectively enantiomer-selective, population transfer 
in Sec.~\ref{sec:strategies} and conclude
in Sec.~\ref{sec:concl}.

\section{Model}
\label{sec:model}

\subsection{Chiral Molecules as Asymmetric Top Rotors}
We consider the interaction of chiral molecules
with electromagnetic radiation,  described by the Hamiltonian 
\begin{equation}
  \hat H^{(\pm)}(t) = \hat H_0 + \hat H_{int}^{(\pm)} (t)\,,
\end{equation}
where the subscript $(\pm)$ denotes the two enantiomers. The molecular Hamiltonian $\hat H_0$ is the same for both enantiomers, except for a very small, parity-violating energy shift which we neglect here.
The dynamics of each enantiomer, induced by the electromagnetic field, is obtained by solving the time-dependent Schr\"odinger equation,
\begin{equation}\label{quantumcontrol}
  \mathrm{i} \hbar \dfrac{d}{dt}|\psi^{(\pm)}(t)\rangle=\left [\hat H_0 + \hat H^{(\pm)}_{int}(t) \right ] |\psi^{(\pm)}(t)\rangle\,.
\end{equation}
Since we consider rotational dynamics of molecules in the electronic and vibrational ground state, there are no dissipative mechanisms relevant on the timescale of the dynamics.
Expectation values for a racemic mixture are obtained via the density operator
$\hat\rho(t)=\frac{1}{2}\sum_\pm\ket{\psi^{(\pm)}(t)}\bra{\psi^{(\pm)}(t)}$.

\begin{figure*}[tb]
	\includegraphics[width=1.0\linewidth]{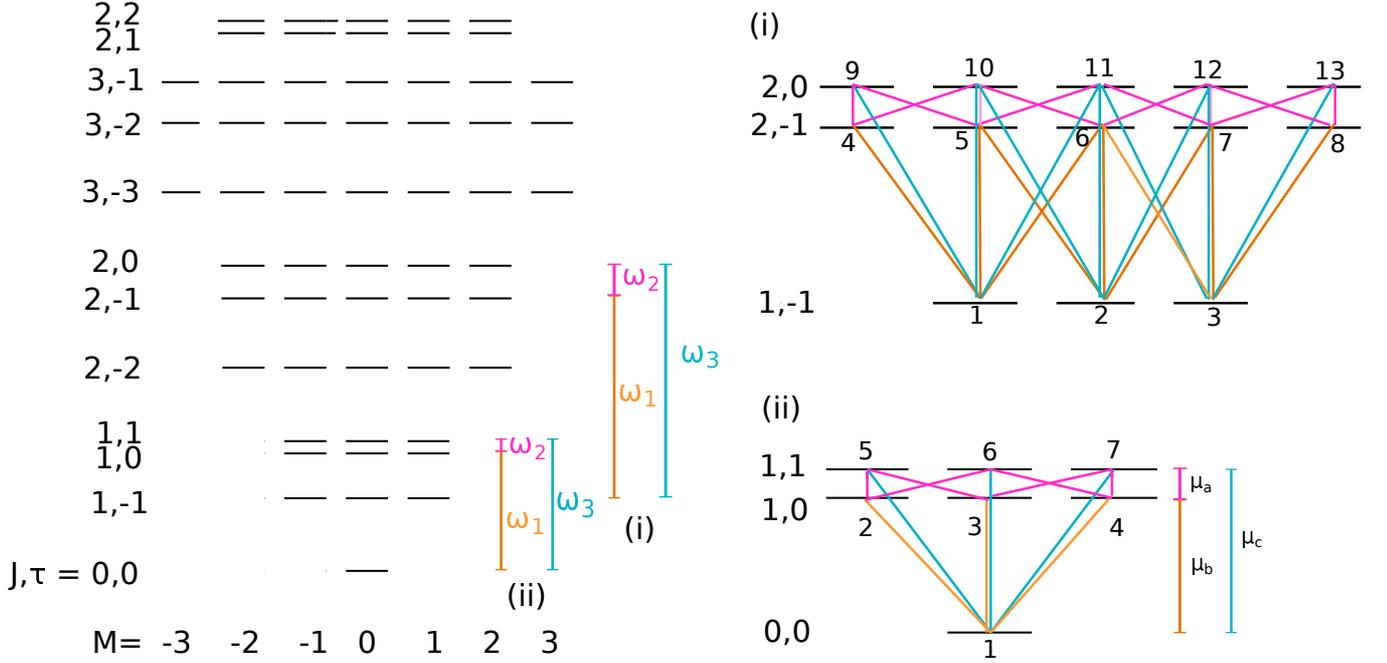} 
	\caption{Left side: Sketch of the energy spectrum of a near-prolate asymmetric top for energies smaller than $E_{2,2}$. The energy levels are denoted by the quantum numbers $J,\tau$. By choosing a set of microwave fields resonant to particular transitions, only a few of the rotational levels are addressed. In (i), the frequencies of the microwave fields are $\omega_1=E_{2,-1}-E_{1,-1}$, $\omega_2=E_{2,0}-E_{2,-1}$, and $\omega_3=E_{2,0}-E_{1,-1}$, in (ii), $\omega_1=E_{1,0}-E_{0,0}$, $\omega_2=E_{1,1}-E_{1,0}$, and $\omega_3=E_{1,1}-E_{0,0}$. Right side: The resulting subsystems of the asymmetric top. The colored lines indicate the transitions induced by $x-$, $y-$, and $z-$polarized fields with frequencies $\omega_1$ (orange), $\omega_2$ (pink), and $\omega_3$ (turquoise).}
	\label{spectrum_asym_top}
\end{figure*}

We assume the molecules to be sufficiently rigid to model them as asymmetric tops. The molecular Hamiltonian $\hat H_0$ then becomes~\cite{KochRMP}
\begin{equation}
  \hat H_0 = \hat H_{rot} =  A {\hat J_a}^2 + B {\hat J_b}^2 + C {\hat J_c}^2\,,
  \label{eq:hrot}
\end{equation}
where $\hat J_a$, $\hat J_b$, and $\hat J_c$ are the angular momentum operators with respect to the principal molecular axes,
and $A > B > C$ are the rotational constants. We adopt the standard approach~\cite{KochRMP} of expressing the asymmetric top eigenstates as superpositions of symmetric top eigenstates $|J,K,M\rangle$,
\begin{equation}
  | J, \tau, M \rangle = \sum_{K} c_{K}^{J} (\tau) |J,K,M\rangle  \,,
  \label{asym_top}
\end{equation}
with symmetric top eigenenergies
\begin{equation}
E^{sym}_{J,K} = B J (J+1) + (A-B) K^2\,,
\label{ev_sym_top}
\end{equation}
where $J$ denotes the rotational quantum number, $J=0, 1,2,\ldots$, and $M$ and $K$ are the projection quantum numbers,
$M=-J, -J+1,\ldots,J$ and $K=-J, -J+1,\ldots,J$, which describe the orientation with respect to a space-fixed and a molecule-fixed axis.  
Note that in Eq.~\eqref{asym_top}, states with different $K$ but the same $J$ and $M$ are mixed. For each $J$, the coefficients $c_{K}^{J}$ and the asymmetric top eigenenergies $E_{J,\tau}$  are obtained by diagonalizing the corresponding $(2J+1)$-dimensional matrix.    
The index $\tau=-J,-J+1,\ldots,J$ counts the asymmetric top states corresponding to a given $J$, starting with the one with lowest energy. 
The spectrum of a near-prolate asymmetric top is sketched in Fig.~\ref{spectrum_asym_top}.

In the electric dipole approximation, the interaction of an asymmetric top with $f$ electric fields linearly polarized along one of the laboratory frame directions can be written as
\begin{equation}
  \hat H_{int}^{(\pm)} = \sum_{i=1}^{f} \hat H_i^{(\pm)} E_i (t) = - \sum_{i=1}^f \hat \mu_i^{(\pm)} E_i(t)\,.
  \label{Hint_def}
\end{equation}
We denote the electric fields by
  ${\vec E}_i(t) = {\vec e}_i E_i(t)$
with amplitude
  $E_i(t) = {\cal E}_i(t) \cos (\omega_i t + \varphi_i)$,
where ${\cal E}_i(t)$ is the envelope and $\omega_i$ and $\varphi_i$ are  frequency and phase of the field. 
The  polarization vector ${\vec e}_i$ can be either ${\vec e}_x$, ${\vec e}_y$ or ${\vec e}_z $.
In Eq.~(\ref{Hint_def}), the dipole moments, given in the laboratory-fixed frame with 
$\hat \mu_i^{(\pm)}$ equal to $\hat \mu_x^{(\pm)}$, $\hat \mu_y^{(\pm)}$, or $\hat \mu_z^{(\pm)} $, 
are connected to the dipole moments $\mu_\sigma^{(\pm)}=( \mu_a^{(\pm)},  \mu_b^{(\pm)},  \mu_c^{(\pm)} )$ in the molecule-fixed frame by a rotation~\cite{Zare88,Leibscher19},
\begin{widetext}
  \begin{eqnarray}
    \hat\mu_x^{(\pm)}  &=& \frac{\mu_a^{(\pm)}}{\sqrt{2}}  \left( D_{-10}^1  - D_{10}^1 \right)  + \frac{\mu_b^{(\pm)}}{ 2} \left ( D_{11}^1 - D_{1-1}^1  - D_{-11}^1 + D_{-1-1}^1\right) -  \mathrm{i} \frac{\mu_c^{(\pm)}}{2} \left ( D_{11}^1 + D_{1-1}^1  - D_{-11}^1 - D_{-1-1}^1\right), \nonumber \\
    \hat\mu_y^{(\pm)}  &=& -\mathrm{i} \frac{\mu_a^{(\pm)}}{\sqrt{2}}  \left( D_{-10}^1  + D_{10}^1 \right)  + \mathrm{i} \frac{\mu_b^{(\pm)}}{ 2} \left ( D_{11}^1 - D_{1-1}^1  + D_{-11}^1 - D_{-1-1}^1\right) + \frac{\mu_c^{(\pm)}}{2} \left ( D_{11}^1 + D_{1-1}^1  + D_{-11}^1 + D_{-1-1}^1\right), \nonumber \\
    \hat \mu_z^{(\pm)}  &=&  \mu_a^{(\pm)} D_{00}^1 - \frac{\mu_b^{(\pm)}}{\sqrt{2} } \left ( D_{01}^1 - D_{0-1}^1 \right) +  \mathrm{i} \frac{\mu_c^{(\pm)}}{\sqrt{2}} \left ( D_{01}^1 + D_{0-1}^1 \right),
                            \label{mu_projection}
  \end{eqnarray}  
\end{widetext}
where $D_{MK}^J$ denote the elements of the Wigner $D$-matrix~\footnote{Note that each element of the Wigner $D$-matrix represents an operator due to its dependence on the Euler angles.}.
For chiral molecules with $C_1$-symmetry, all three components $\mu_\sigma^{(\pm)}$ are non-zero. Moreover, $|\mu_\sigma^{(+)}|=|\mu_\sigma^{(-)}|$ and
\begin{equation}
  \mu_a^{(+)} \mu_b^{(+)}  \mu_c^{(+)} = - \mu_a^{(-)} \mu_b^{(-)}  \mu_c^{(-)},
  \label{dipole_sign}
\end{equation}
i.e., the two enantiomers differ in the sign of one of the Cartesian components of the dipole moment. Equation~\eqref{dipole_sign} is the basis of enantiomer-specific three-wave mixing \cite{PattersonNature13}.
    
In the asymmetric top eigenbasis Eq.(\ref{asym_top}), the interaction Hamiltonian 
contains matrix elements of the form
\begin{widetext}
    \begin{eqnarray}
    \langle J'', \tau'', M'' | D_{MK}^1 | J', \tau', M' \rangle = \sum_{K',K''} 
    c_{K'}^{J'} \left ( c_{K''}^{J''} \right )^\ast \langle J'', K'', M'' |D_{MK}^1|J', K', M' \rangle.
    \label{transition_asym}
    \end{eqnarray}
    with 
    \begin{eqnarray}
    \langle J'', K'', M'' | D^1_{MK} | J', K', M' \rangle &=& \sqrt{2 J'' +1} \sqrt{2 J' +1} (-1)^{M''+K''}  
    \times  \left(    \begin{array}{ccc}  J' & 1& J'' \\ M' & M & -M'' \end{array} \right)
    \left( \begin{array}{ccc} J' & 1 & J'' \\ K' & K & -K'' \end{array} \right).
    \label{w3j}
    \end{eqnarray}
\end{widetext}
The Wigner 3j-symbols in Eq.~\eqref{w3j} determine the selection rules, namely $J''-J' = 0, \pm 1$ and $K'' = K' + K$ as well as $M'' = M' + M$ where the value of $M$ is determined by the electric field polarization in Eq.~(\ref{transition_asym}) with $M =0$ for $z$-polarized and  $M = \pm 1$ for $x$- or $y$-polarized fields.

\subsection{Control problem}

Our goal is to transfer population which is initially distributed over a degenerate manifold into quantum states which are energetically separated. Such a transfer can serve as  precursor for distilling population out of an incoherent mixture.

For a racemic mixture of chiral molecules, the two enantiomers initially occupy the same rotational states since they possess the same rotational spectrum. The initial state is thus described by the density matrix
\begin{equation}\label{eq:rho-racemic}
\rho(t=0) = \frac{1}{2} \left(\rho^{(+)}(t=0) + \rho^{(-)}(t=0)\right).
\end{equation}
At non-zero temperatures,
the state of each enantiomer is given by a thermal ensemble, 
\begin{equation}\label{eq:rho-ini-pm}
\rho^{(\pm)} (t=0) = \sum_{J_0,\tau_0,M_0} p_{J_0,\tau_0}|J_0,\tau_0,M_0 \rangle \langle J_0,\tau_0,M_0 |,
\end{equation}
where $p_{J_0,\tau_0}$ is the Boltzmann weight of the rotational level denoted by $J_0$ and $\tau_0$ and the incoherent sum over the degenerate $M_0$-states accounts for the isotropic angular distribution of molecules in the gas phase.

We seek to achieve the population transfer with narrow-bandwidth pulses such that only resonant transitions (with  $E_{J'',\tau''} -  E_{J',\tau'} = \hbar \omega_i$) need to be considered. This assumption is typically well-justified in the microwave regime and reduces the number of non-zero matrix elements in the interaction Hamiltonian to those 
with the appropriate combination of frequency $\omega_i$ and electric field polarization ${\vec e}_i$. Moreover,
for $f$ combinations of polarization and frequency, we obtain 
$f$ linearly independent interaction matrices ${\bf H}_i^{(\pm)}$,
expressing the interaction Hamiltonian~\eqref{Hint_def} in the basis of the asymmetric top states. Since a given set of microwave fields
addresses only a finite number of rotational levels,
we can describe the dynamics in a comparatively small rotational subsystem.
Figure~\ref{spectrum_asym_top} shows two examples of such subsystems that are relevant for microwave three-wave mixing in chiral molecules: (i) Fields with frequencies  $\omega_1=E_{2,-1}-E_{1,-1}$, $\omega_2=E_{2,0}-E_{2,-1}$, and $\omega_3=E_{2,0}-E_{1,-1}$ couple the states with rotational energies $E_{1,-1}$, $E_{2,-1}$, and $E_{2,0}$, while in Fig.~\ref{spectrum_asym_top}(ii) the states with $E_{0,0}$, $E_{1,0}$, and $E_{1,1}$ are addressed.   

At zero temperature, when 
$\rho^{(\pm)} = |0,0,0 \rangle \langle 0,0,0 |$, excitation with three microwave pulses with $x$-, $y$-, and $z$-polarization and frequencies $\omega_1$, $\omega_2$, and $\omega_3$, as shown in Fig.~\ref{spectrum_asym_top}(ii), is predicted to lead to 100$\%$ enantio-selective excitation \cite{Leibscher19,LehmannJCP18}.  However, microwave three-wave mixing experiments typically address rotational states with $J=1$ and $J=2$, as shown in Fig.~\ref{spectrum_asym_top}(i), or with $J=2$ and $J=3$~\cite{ShubertAngewandte14,PerezAngewandte17,EibenbergerPRL17}  where all levels with given $J_0$, $\tau_0$ are $(2J_0+1)$-fold degenerate.
This degeneracy results in incomplete enantio-selective population transfer in state-of-the-art three-wave mixing~\cite{PattersonNature13,ShubertAngewandte14,LobsigerJPCL15,PerezAngewandte17,EibenbergerPRL17}, even if temperature effects are not considered. Below we will show that complete enantio-selective excitation into energetically separated quantum states can be achieved in a racemic mixture, cf. Eq.~\eqref{eq:rho-racemic}, 
despite the $M$-degeneracy of the rotational states. In passing, we furthermore show that population distributed over degenerate levels in Eq.~\eqref{eq:rho-ini-pm} can also be energetically separated.

\section{Theoretical framework for controllability analysis}
\label{sec:Lie}

Given the model of a quantum system and its interaction with external fields,
the controllability analysis consists in addressing the question of whether  a control target can or cannot be reached. 
This is in contrast to control synthesis which devises the shapes of the external fields that drive the system to the target in the best possible way~\cite{GlaserEPJD15}. Controllability is thus a prerequisite for control synthesis. 

Controllability may refer to a single quantum system or an ensemble of quantum systems that shall be controlled simultaneously with only a few control fields~\cite{GlaserEPJD15}. Here, we adapt the notion of simultaneous controllability to the specific task of enantiomer-selective population transfer. In Sec.~\ref{subsec:controllability},
we first recall the basic mathematical concepts for controllability analysis,  before defining enantio-selective controllability in Sec.~\ref{subsec:enantiosel}.

\subsection{Lie rank condition and spectral gap excitation}
\label{subsec:controllability} 

A quantum system is said to be completely controllable if we can steer it 
from any initial state to any final state 
by suitably choosing the (possibly time-dependent) shapes of external fields.
If a system is completely controllable, any incoherent initial state 
can be steered to any final state with the same purity 
\cite{Alessandro2008}.

For finite dimensional systems, a sufficient condition for a system to be completely controllable is requiring the Lie algebra of its Hamiltonian to be of full rank~\cite{Alessandro2008},
\begin{equation}\label{bracket}
\dim\Big(\mathrm{Lie}\left\{\mathrm{i}{\bf H}_0,\ldots,\mathrm{i} {\bf H}_f\right\} \Big)=N^2-1 \,
\end{equation}
where $N$ denotes the Hilbert space dimension and $\mathrm{Lie}\{\mathrm{i}{\bf H}_0,\ldots,\mathrm{i} {\bf H}_f\}$ the maximal real vector space of matrices obtained by repeatedly taking commutators (Lie brackets) between the elements of the total Hamiltonian.
We consider here, without loss of generality, traceless Hamiltonians~\footnote{Otherwise the dimension in Eq.~\eqref{bracket} should be $N^2$ for a system to be controllable.}. A quantum system is not completely controllable if it possesses a symmetry. The existence of a symmetry operator, i.e., an operator that commutes with the Hamiltonian, is equivalent to the existence of a conserved quantity. As a result, the Hamiltonian can be written in block-diagonal form, without transition matrix elements connecting the blocks such that the system can be controlled only within the symmetry-enforced subsystems. 

The Lie rank condition \eqref{bracket} can be effectively checked on reduced Hamiltonians,
obtained by considering only transitions resonant to a selected frequency. 
More precisely, denoting the eigenstates and eigenvalues of ${\hat H}_0$
by $|\psi_1\rangle,\ldots,|\psi_N\rangle$ and $\lambda_1,\ldots,\lambda_N$, and the set of spectral gaps of the system 
by $\Sigma=\{|\lambda_i-\lambda_j|,\;i,j=1,\dots,N\}$, 
one considers frequencies $\omega\in \Sigma$ and  matrices ${\bf H}_{\omega,a}$ defined by 
\begin{equation}\label{filtred}
\langle \psi_h|\hat H_{\omega,a}|\psi_k\rangle = \begin{cases}
\langle \psi_h|\hat H_{a}|\psi_k\rangle & \text { if } |\lambda_h-\lambda_k|=\omega \\
0 & \text{ otherwise}.
\end{cases}
\end{equation}
Then, if one can find frequencies $\{\omega_1,\dots,\omega_k\}\subset \Sigma$ such that 
\begin{widetext}
	\begin{equation}\label{spectralbracket}
	\dim\Big(\mathrm{Lie}\left\{\mathrm{i}{\bf H}_0,\mathrm{i} {\bf H}_{\omega_{i},a} \mid a\in \left\{1,\ldots,f\right\}, 
	i\in \left\{1,\ldots,k\right\} \right\}\Big)=N^2-1\,,
	\end{equation}  
\end{widetext}
the system is controllable~\cite{BCS}. Equations \eqref{bracket} and \eqref{spectralbracket} are equivalent sufficient conditions for controllability~\cite{chambrion,BCS}.
Equation~\eqref{spectralbracket} implies that the Lie algebra generated by 
${\bf H}_0$ and the various ${\bf H}_{\omega_i,a}$ is all of $\mathfrak{su}(N)$, i.e.,  the Lie algebra of traceless  $N\times N$ skew-Hermitian matrices.

The conditions for controllability, Eqs. \eqref{bracket} and \eqref{spectralbracket}, hold for a finite-dimensional system whereas the spectrum of a quantum rotor is, in principle, infinite-dimensional.
Equation \eqref{spectralbracket} can be used to check the approximate controllability of a finite-dimensional subsystem of a system with an infinite number of energy levels. To be able to do it, 
all frequencies $\omega \in \Sigma$ that are required for controllability of the subsystem must be different from the frequencies coupling two states, one in the subsystem and the other one outside.
Approximate controllability then means that 
each target of the subsystem can be reached  with arbitrary precision, provided that the time is sufficiently large.    
This is based on the fact that, if a frequency $\omega\in\Sigma$ is resonant with a finite number of spectral gaps only, the operators ${\bf H}_{\omega,a}$ do not address transitions in the total rotational state space, but only within a finite-dimensional part of it.

Truncating the infinite-dimensional Hilbert space by a finite-dimensional subspace is equivalent to a Galerkin approximation \cite{CMSB}. Introducing,  for $n=1,2,\ldots$,
$\Sigma_n=\{|\lambda_i-\lambda_j|,\;i,j=1,\dots,n\}$ as the 
set of spectral gaps of the 
$n$-th approximation of ${\bf H}_0$ and 
$\hat \Sigma_n=\{|\lambda_i-\lambda_j|,\;i=1,\dots,n,\;j=n+1,n+2,\dots\}$, and defining $\Xi_n=\{\omega\in\Sigma_n \mid \omega\neq 0,\; \omega \notin \hat\Sigma_n\}$ which  contains
exactly the frequencies in $\Sigma_n$ non-resonant with higher approximations of the spectrum, the following can be stated: If, for any $n_0$ one can find an $n>n_0$ such that
\begin{widetext}
	\begin{equation}\label{LGC}
	\dim\Big(\mathrm{Lie}\left\{\mathrm{i}{\bf H}_0,\mathrm{i} {\bf H}_{\omega,a} \mid a\in \left\{1,\ldots,f\right\},
	\omega\in \Xi_n \right\}\Big)=n^2-1\,,
	\end{equation}  
\end{widetext}
 then the system is approximately controllable, i.e., it is possible to steer it from any initial state arbitrarily close to any final state~\cite{BCS}.
Equation \eqref{LGC}, called Lie--Galerkin condition, is a sufficient condition for approximate controllability of infinite-dimensional systems.
Moreover, if the Lie--Galerkin condition holds, the finite-dimensional projections are exactly controllable, that is, 
one can find a time $T$ such that the finite-dimensional projections of the infinite-dimensional propagator are exactly the finite-dimensional projections of the target propagator~\cite{CS}. 

\subsection{Enantio-selective controllability} 
\label{subsec:enantiosel}
For a rigid asymmetric top, we can apply the controllability analysis according to Eq.~\eqref{spectralbracket} by identifying 
the matrices ${\bf H}_{\omega_i,a}$ with the $f$ linearly independent interaction matrices ${\bf H}_i^{(\pm)}$.
If such a molecule, evolving according to Eq.~\eqref{quantumcontrol}, is  controllable, one can --- at least in principle --- find electric fields which steer a given initial state, $|\psi^{(+)}(t=0) \rangle$ or $\rho^{(+)}(0)$,  to a desired target state, $|\psi_ {target}^{(+)}\rangle$ 
or $\rho^{(+)}_{target}$ (with same purity).
However, controllability of Eq.(\ref{quantumcontrol}) does not imply that one can,  with the same set of control fields, steer $|\psi^{(+)}(0) \rangle$  to  $|\psi_ {final}^{(+)}\rangle$ and
$|\psi^{(-)}(0) \rangle$  to  $|\psi_ {final}^{(-)}\rangle$ simultaneously. To capture such a control target, we introduce the concept of enantio-selective controllability.
It corresponds to the problem of simultaneously controlling two evolutions,  i.e., the evolution of the two enantiomers, governed by the same molecular Hamiltonian ${\hat H}_0$ and controlled with the same fields $E_i(t)$. 

We call a system enantio-selective controllable if both enantiomers are individually controllable with the same set of external fields. 
To analyze enantio-selective controllability, we construct a composite system, defined on a Hilbert space which is the tensor sum  $\mathcal{H}\oplus \mathcal{H}$ of the (identical) rotational state spaces of the two enantiomers.
The corresponding Hamiltonian is block-diagonal,
\begin{eqnarray} 
{\bf H}^{chiral}(t) &=& {\bf H}_0^{chiral} + {\bf H}_{int}^{chiral}(t) \nonumber \\
&=& \left( \begin{array}{cc}
{\bf H}_{0} & 0  \\
0 & {\bf H}_{0}  \end{array} \right) 
+         \left( \begin{array}{cc}
{\bf H}_{int}^{(+)}(t) & 0  \\
0 & {\bf H}_{int}^{(-)}(t)  \end{array} \right) \nonumber \\ 
 &=&\;{\bf H}_{0}\;\oplus\; {\bf H}_{0} \quad+\quad  {\bf H}_{int}^{(+)}(t)\; \oplus\; {\bf H}_{int}^{(-)}(t) \nonumber \\
\label{Ham_int_tot}
\end{eqnarray}  
with ${\bf H}_{0}$ and  ${\bf H}_{int}^{(\pm)}(t)$ being the matrix representations of $\hat H_0$ and $\hat H_{int}^{(\pm)}$ in the asymmetric top eigenbasis, Eq.~\eqref{asym_top}~\footnote{The block-diagonal structure of the Hamiltonian in Eq.~\eqref{Ham_int_tot} is a result of parity conservation in a rigid rotor, i.e., within this description enantiomers cannot be converted into each other.}.
A system described by a block-diagonal matrix with two blocks of the size $N \times N$ is 
controllable if its Lie algebra has the dimension $2(N^2-1)$. In other words, due to the block structure of Eq.~\eqref{Ham_int_tot}, the system is enantio-selective controllable if its Lie algebra is $\mathfrak{su}(N)\oplus \mathfrak{su}(N)$. This corresponds exactly to the sufficient condition for simultaneous controllability, see e.g. Ref.~\cite{dirr,turinici}. The Lie rank condition for enantio-selective controllability is equivalent to saying that one is able to reproduce (by taking enough commutators) any operator of the form ${\bf A}\oplus {\bf 0}$ and ${\bf 0}\oplus {\bf B}$  $\forall\;A,B\in\mathfrak{su}(N)$: ${\bf A}\oplus {\bf 0}$ steers any initial state to any final state within the first enantiomer leaving the state of the second enantiomer unchanged, while ${\bf 0}\oplus {\bf B}$ steers any initial state to any final state within the second enantiomer without changing the state of the first enantiomer.

\section{Controllability of asymmetric quantum rotors}
\label{sec:controllabilityresults}

We now use the concepts summarized in Sec.~\ref{sec:Lie} to analyze controllability and enantio-selective controllability of rigid quantum rotors. 
Generally, controllability of quantum rotors is difficult to prove 
due the presence of the $M$- (and for symmetric tops $K$-) degeneracies, which lead to linear combinations of elements of the Lie algebra which are seemingly not linearly independent. 
For infinite-dimensional linear tops and for infinite-dimensional accidentally symmetric tops, the problem of $M$-degeneracy has recently been overcome using the  Lie--Galerkin control technique \cite{BCS,Boscain19}. 
One can prove that accidentally symmetric tops are, however, not enantio-selective controllable since their $K$-degeneracy prevents the simultaneous controllability of the two enantiomers. 
The proof of enantio-selective controllability for the complete spectrum of an asymmetric top is an ongoing challenge. Here, we exploit the fact that microwave three-wave mixing spectroscopy relies on 
resonant excitations confining (up to an arbitrarily small error) the rotational dynamics to a subsystem with finitely many rotational levels,
cf.  Fig.~\ref{spectrum_asym_top}. For our purposes it is thus sufficient to analyze the controllability and enantio-selective controllability of specific subsystems of a chiral asymmetric top. To carry out this analysis, we introduce generalized Pauli matrices (Sec.~\ref{subsec:generalpauli}) and apply them to the enantio-selective controllability of specific rotational subsystems (Sec.~\ref{subsec:subsystems} and \ref{subsec:subsets}).

\subsection{Generalized Pauli matrices}
\label{subsec:generalpauli}
To analyze controllability of a finite-dimensional rotational subsystem described by ${\bf H}_0$ interacting with a set of $f$ electromagnetic fields via the interaction Hamiltonians $i {\bf H}_{\omega_i,a}$, we express $i {\bf H}_{\omega_i,a}$
as linear combinations of the generalized Paul matrices \cite{Boscain19},
\begin{eqnarray}\label{basis}
{\bf G}_{j,k}&=&e_{j,k}-e_{k,j}\,, \nonumber \\
{\bf F}_{j,k}&=&\mathrm{i}e_{j,k}+\mathrm{i}e_{k,j}\,, \nonumber \\
{\bf D}_{j,k}&=&\mathrm{i}e_{j,j}-\mathrm{i}e_{k,k}\,,
\end{eqnarray}
where $e_{j,k}$ is the matrix whose entries are all zero except for the entry in row $j$ and column $k$ which is equal to $1$. These operators (with $j,k=1,\ldots,n$) form a basis of the Lie algebra 
$\mathfrak{su}(n)$. 
Thus, we need to show that we obtain elements of the Lie algebra which are proportional to each of the operators ${\bf G}_{j,k}$, ${\bf F}_{j,k}$, and  ${\bf D}_{j,k}$ alone by repeatedly taking commutators between  $i {\bf H}_{\omega_i,a}$ and  $i {\bf H}_0$.
It is useful to recall the commutator relations between the generalized Paul matrices, 
\begin{subequations}\label{eq:commutators}
\begin{eqnarray}
\left[ {\bf G}_{j,k},{\bf G}_{k,n} \right]&=& {\bf G}_{j,n}\,, \nonumber \\
\left[ {\bf F}_{j,k},{\bf F}_{k,n} \right] &=& -{\bf G}_{j,n}\,, \nonumber \\
\left[ {\bf G}_{j,k},{\bf F}_{k,n} \right] &=& {\bf F}_{j,n} \,, 
\label{relation1} 
\end{eqnarray}
and 
\begin{eqnarray}
\label{relation2}
\left[{\bf G}_{j,k},{\bf F}_{j,k}\right] &=&2{\bf D}_{j,k} \,,\nonumber \\ 
\left[{\bf F}_{j,k},{\bf D}_{j,k}\right] &=& 2{\bf G}_{j,k}\,.
\end{eqnarray}
Moreover,  operators which couple disjunct pairs of states commute,   
\begin{equation}\label{relation3}
[{\bf T}_{j,k},{\bf U}_{j',k'}]=0 \quad \text{ if } \{j,k\}\cap  \{j',k'\}=\emptyset ,
\end{equation}
with ${\bf T},{\bf U}\in \{{\bf G},{\bf F},{\bf D}\}$.
Finally, the commutators with the rotational Hamiltonian are given by
\begin{eqnarray}\label{relation4}
\left[\mathrm{i}{\bf H}_0,{\bf G}_{j,k} \right] &=&-\Delta E_{k,j} {\bf F}_{j,k}\,, \nonumber \\ 
\left[\mathrm{i} {\bf H}_0,{\bf F}_{j,k}\right] &=& \Delta E_{k,j} {\bf G}_{j,k}\,.
\end{eqnarray}
\end{subequations}
where $\Delta E_{k,j}$ is the energy gap between states $j$ and $k$.

\subsection{Complete controllability and enantio-selective controllability of rotational subsystems of the type $J/J+1/J+1$}
\label{subsec:subsystems}

In order to analyze controllability and enantio-selective controllability in the subsystem made up of rotational states with energies $E_{J,\tau}$, $E_{J+1,\tau'}$, and $E_{J+1,\tau''}$, cf. Fig.~\ref{spectrum_asym_top}, we diagonalize the asymmetric top Hamiltonian $\hat{H}_0=\hat{H}_{rot}$ for this particular subsystem and compute the dimension of the Lie algebra generated by a set of control fields. Controllability then has to be proven individually for each subsystem of interest.

The proof involves two steps. First, we prove complete controllability of the rotational subsystem of a single enantiomer, made up by 
all rotational states with energies $E_{J,\tau}$, $E_{J+1,\tau'}$, and $E_{J+1,\tau''}$. This result by itself is already quite remarkable. It implies that each level, including the degenerate ones, can be addressed separately with electric fields alone, and it is not necessary to lift the degeneracy, for example with a magnetic field. To carry out this part of the proof,
we need to consider four control fields with linear polarization directions $p_i$ and frequencies $\omega_1$ and $\omega_2$ as defined in Fig.~\ref{spectrum_asym_top}, chosen
such as to 
induce transitions via the dipole moments $\mu_b$ and $\mu_a$, respectively. The corresponding interaction Hamiltonians ${\bf H}_{\omega_1,p_1}$, ${\bf H}_{\omega_1,p_2} $, ${\bf H}_{\omega_2,p_3}$, and ${\bf H}_{\omega_2,p_4} $ are expressed in terms of the generalized Pauli matrices (\ref{basis}).
We analyze the resulting Lie algebra by repeatedly taking commutators. Since the dimension of the subsystems is $l_J=(2J+1)+2(2J+3)$, the Lie algebra has to be $\mathfrak{su}(l_J)$ for the subsystem to be controllable.

In a second step, we prove enantio-selective controllability by adding a control field with frequency $\omega_3=\omega_1+\omega_2$ and  interaction Hamiltonian ${\bf H}_{\omega_3,p_5}$. As indicated in Fig.\ref{spectrum_asym_top}, such a field couples rotational states via the dipole moment $\mu_c$. The corresponding Lie algebra has to be
$\mathfrak{su}(l_J) \oplus \mathfrak{su}(l_J) $ for the subsystem to be enantio-selective controllable.
In the following, we demonstrate controllability of the $J=0/1/1$-subsystem, example (ii) in Fig.~\ref{spectrum_asym_top}. Another example, a subsystem containing rotational states with $J=2/3/3$, is treated in the Appendix.
The extension of both steps of the proof to $J/J+1/J+1$ subsystems with $J>2$ is tedious but straightforward. 
Note  in particular that the four, respectively five, different fields are sufficient to prove complete controllability, respectively enantio-selective controllability, independently of the specific choice of $J$. 

To prove controllability of the $J=0/1/1$-subsystem of a single enantiomer, we write the rotational Hamiltonian,
\begin{equation}
{\bf H}_{0} =\mathrm{diag}\left(E_{0,0},E_{1,0},E_{1,0},E_{1,0},E_{1,1},E_{1,1},E_{1,1} \right) \nonumber
\end{equation}
and consider a set of four interaction operators,
\begin{equation}
\mathcal{X}_1=\{\mathrm{i}{\bf H}_{\omega_1,x},\mathrm{i}{\bf H}_{\omega_1,z},\mathrm{i}{\bf H}_{\omega_2,y},\mathrm{i}{\bf H}_{\omega_2,z}\}\,.
\end{equation}
Here, we have chosen the polarization directions to be $p_1=x$, $p_2=z$,
$p_3=y$, and $p_4=z$.
We have to show that 
\begin{equation}\label{su(7)}
\mathrm{Lie}\{ \{ \mathrm{i}{\bf H}_{0}\}\cup\mathcal{X}_1 \}\}=\mathfrak{su}(7),
\end{equation}
since the Hilbert space $\mathcal{H}^{(\pm)}$ coincides with $\mathbb{C}^7$. 
Using Eqs.~\eqref{mu_projection} and \eqref{w3j},
we can write the interaction operators as
linear combinations of the generalized Pauli matrices, 
\begin{eqnarray}
	\mathrm{i}{\bf H}_{\omega_1,x}&=&\frac{\mu_b}{\sqrt{6}}({\bf G}_{1,4}-{\bf G}_{1,2})\,, \nonumber \\
	\mathrm{i}{\bf H}_{\omega_1,z}&=&-\frac{\mu_b}{\sqrt{3}} {\bf G}_{1,3}\,, \nonumber \\
	\mathrm{i}{\bf H}_{\omega_2,y}&=&\frac{\mu_a}{2\sqrt{2}}({\bf G}_{3,5}+{\bf G}_{4,6}-{\bf G}_{2,6}-{\bf G}_{3,7})\,, \nonumber \\
	\mathrm{i}{\bf H}_{\omega_2,z}&=&\frac{\mu_a}{2}(-{\bf F}_{2,5}+{\bf F}_{4,7})\,,
	\label{Hint_normal}
\end{eqnarray}
where the matrix elements are labeled according to Fig.~\ref{spectrum_asym_top}(ii). 
For example, $\mathrm{i}{\bf H}_{\omega_1,x}=\mu_b({\bf G}_{1,4}-{\bf G}_{1,2})$ means that the field with $x$-polarization and frequency $\omega_1$ couples the state $1=\ket{0,0,0}$ to the states $4=\ket{1,0,1}$ and $2=\ket{1,0,\minusone}$. 
With the commutator relations~\eqref{eq:commutators}, we find 
\begin{eqnarray*}
	[\mathrm{i}{\bf H}_{0},\mathrm{i}{\bf H}_{\omega_2,z}]\propto -{\bf G}_{2,5}+{\bf G}_{4,7}=:J(\mathrm{i}{\bf H}_{\omega_2,z})
\end{eqnarray*}
and 
\begin{eqnarray}\label{commutators2}
\left[\mathrm{i}{\bf H}_{\omega_1,x},J(\mathrm{i}{\bf H}_{\omega_2,z})\right]&\propto&{\bf G}_{1,5}+{\bf G}_{1,7}, \nonumber \\
\left[\mathrm{i}{\bf H}_{\omega_1,z},\mathrm{i}{\bf H}_{\omega_2,y}\right]&\propto&{\bf G}_{1,5}-{\bf G}_{1,7}.
\end{eqnarray}
Taking the sum and the difference, we obtain
\begin{eqnarray}
	\left[\mathrm{i}{\bf H}_{\omega_1,x},J(\mathrm{i}{\bf H}_{\omega_2,z})\right]+\left[\mathrm{i}{\bf H}_{\omega_1,z},\mathrm{i}{\bf H}_{\omega_2,y}\right]&\propto& {\bf G}_{1,5} \nonumber \\
	 \left[\mathrm{i}{\bf H}_{\omega_1,x},J(\mathrm{i}{\bf H}_{\omega_2,z})\right]-\left[\mathrm{i}{\bf H}_{\omega_1,z},\mathrm{i}{\bf H}_{\omega_2,y}\right]&\propto& {\bf G}_{1,7}.
\end{eqnarray}
In this way, we generate operators that separately address the transitions $1\leftrightarrow 5$ and $1\leftrightarrow 7$, i.e., that act separately on two degenerate $M$-states. Moreover, we find 
\begin{eqnarray}
	\left[{\bf G}_{1,7},J(\mathrm{i}{\bf H}_{\omega_2,z})\right] &\propto& {\bf G}_{1,4} \nonumber \\
	\left[{\bf G}_{1,5},J(\mathrm{i}{\bf H}_{\omega_2,z})\right] &\propto& {\bf G}_{1,2}\nonumber \\
	\left[{\bf G}_{1,2},\mathrm{i}{\bf H}_{\omega_2,y}\right]&\propto& {\bf G}_{1,6}.
 \end{eqnarray}
So far, we have obtained all elements ${\bf G}_{j,k}$ with $j=1$.   
Applying Eq.~\eqref{relation1} to these elements, we get all remaining elements ${\bf G}_{j,k}$, $j,k\in\{1,\dots,7\}$, and using Eqs.~\eqref{relation4} and \eqref{relation2}, we obtain all elements ${\bf F}_{j,k}$ and ${\bf D}_{j,k}$, $j,k\in\{1,\dots,7\}$. Since the elements ${\bf G}_{j,k},{\bf F}_{j,k},{\bf D}_{j,k}$ form a basis of $\mathfrak{su}(7)$, we have proven that the Lie algebra is $\mathfrak{su}(7)$. The subsystem is thus controllable with the set of control fields $\mathcal{X}_1$. In the same way, it can also be shown that the system is not controllable if any of the four fields contained in
$\mathcal{X}_1$ is left out. 

In the second step, we extend the proof to the composite system of both enantiomers, showing enantio-selective controllability. Without loss of generality, we assume that the dipole moments of the two enantiomers are $(\mu_a^{(+)},  \mu_b^{(+)},\mu_c^{(+)} )=( \mu_a, \mu_b,\mu_c)$ and $( \mu_a^{(-)}, \mu_b^{(-)},\mu_c^{(-)}) = (\mu_a,\mu_b, -\mu_c)$. For the interaction Hamiltonians, it follows that
 ${\bf H}^{(+)}_{\omega_1,p_i}= {\bf H}^{(-)}_{\omega_1,p_i}$ and ${\bf H}^{(+)}_{\omega_2,p_i}= {\bf H}^{(-)}_{\omega_2,p_i}$ since, according to Eq.~(\ref{Hint_normal}), these matrices are proportional to $\mu_a$ and $\mu_b$. Thus the four fields contained in $\mathcal{X}_1$ applied to the composite system result in
\begin{equation}\label{firststep}
\begin{split}
\mathrm{Lie}\{ \{ \mathrm{i}{\bf H}_{0}^{chiral}\}&\cup \mathcal{X}_1 \} \\ &= 
\left\{\begin{pmatrix}
A & 0\\
0 & A
\end{pmatrix} \mid A\in \mathfrak{su}(7) \right\}
\end{split}
\end{equation}
as matrices acting on the vector space $\mathcal{H}^{(+)}\oplus\mathcal{H}^{(-)} = \mathbb{C}^{7}\oplus \mathbb{C}^{7}$. For the Lie algebra to be 
$\mathfrak{su}(7) \oplus \mathfrak{su}(7)$, an additional control field with frequency $\omega_3$ is required which leads to the interaction operator
\begin{eqnarray}
\mathrm{i}{\bf H}_{\omega_3,x}^{chiral}= \left( 
\begin{array}{cc}
\mathrm{i}{\bf H}_{\omega_3,x} & 0  \\
0 & -\mathrm{i}{\bf H}_{\omega_3,x}  \end{array} \right) 
\end{eqnarray}
with $ \mathrm{i}{\bf H}_{\omega_3,x}=\frac{\mu_c}{\sqrt{6}}(F_{1,5}-F_{1,7})$ and 
the minus sign in the lower block occuring because of $\mu_c^{(+)}=-\mu_c^{(-)}$. To prove that the system is enantio-selective controllable with the set of five control fields generating the interaction operators
$$\mathcal{X}=\{\mathrm{i}{\bf H}_{\omega_1,x}^{chiral},\mathrm{i}{\bf H}_{\omega_1,z}^{chiral},\mathrm{i}{\bf H}_{\omega_2,y}^{chiral},\mathrm{i}{\bf H}_{\omega_2,z}^{chiral},\mathrm{i}{\bf H}_{\omega_3,x}^{chiral}\}\,,$$
we need to show that 
\begin{widetext}
	\begin{equation}
	\label{enantioselection}
	\mathrm{Lie}\{ \{ \mathrm{i}{\bf H}_{0}^{chiral}\}\cup\mathcal{X}\}=\mathrm{span}\left\{ \begin{pmatrix}
	A & 0\\
	0 & 0
	\end{pmatrix}, \begin{pmatrix}
	0 & 0\\
	0 & A
	\end{pmatrix} \mid A\in \mathfrak{su}(7)\right \},  
	\end{equation}
		since
	$$\mathrm{span}\left\{ \begin{pmatrix}
	A & 0\\
	0 & 0
	\end{pmatrix}, \begin{pmatrix}
	0 & 0\\
	0 & A
	\end{pmatrix} \mid A\in \mathfrak{su}(7)\right \}\cong\mathfrak{su}(7)\oplus \mathfrak{su}(7)\,. $$
\end{widetext}	
    To do so, we consider the matrix
	\[
	{\bf V}:=\begin{pmatrix}
	{\bf G}_{1,5}-{\bf G}_{1,7} & 0\\
	0 & {\bf G}_{1,5}-{\bf G}_{1,7}
	\end{pmatrix}
	\]
which is an element of the Lie algebra generated from the four fields contained in 
$\mathcal{X}_1$, see Eq.~(\ref{commutators2}). Moreover,	
\[
[\mathrm{i}{\bf H}_{0}^{chiral}, \mathrm{i}{\bf H}_{\omega_3,x}^{chiral}]  \propto J(\mathrm{i}{\bf H}_{\omega_3,x}^{chiral})
\]	
with
\[
J(\mathrm{i}{\bf H}_{\omega_3,x}^{chiral}):=\begin{pmatrix}
{\bf G}_{1,5}-{\bf G}_{1,7} & 0\\
0 & -{\bf G}_{1,5}+{\bf G}_{1,7}
\end{pmatrix}.
\]	
We see that ${\bf V}$ and $J(\mathrm{i}{\bf H}_{\omega_3,x}^{chiral})$ differ by the sign of the matrix elements belonging to the second enantiomer. Taking the sum and difference
of the two matrices, we obtain
	\[
	\frac{1}{2}\left(J(\mathrm{i}{\bf H}_{\omega_3,x}^{chiral})+{\bf V}\right)=\begin{pmatrix}
	{\bf G}_{1,5}-{\bf G}_{1,7} & 0\\
	0 & 0
	\end{pmatrix}
	\] 
	and
	\[
	 \frac{1}{2}\left( J(\mathrm{i}{\bf H}_{\omega_3,x}^{chiral})-{\bf V}\right)=\begin{pmatrix}
	0 & 0\\
	0 & -{\bf G}_{1,5}+{\bf G}_{1,7}
	\end{pmatrix} \,,
	\]
which are two operators belonging to the Lie algebras acting only on the first and the second enantiomer, respectively. 
Furthermore,
\begin{widetext}
\begin{equation}\label{sum1}	
\left[\frac{1}{2}\left(J(\mathrm{i}{\bf H}_{\omega_3,x}^{chiral})+{\bf V}\right),\begin{pmatrix}
{\bf G}_{5,7} & 0\\
0 & {\bf G}_{5,7}
\end{pmatrix} \right]
=\left[\begin{pmatrix}
{\bf G}_{1,5}-{\bf G}_{1,7} & 0\\
0 & 0
\end{pmatrix},\begin{pmatrix}
{\bf G}_{5,7} & 0\\
0 & {\bf G}_{5,7}
\end{pmatrix} \right]=\begin{pmatrix}
{\bf G}_{1,7}+{\bf G}_{1,5} & 0\\
0 & 0
\end{pmatrix}
\end{equation}
and finally the sum
\begin{equation}\label{sum2}
\left[\frac{1}{2}\left(J(\mathrm{i}{\bf H}_{\omega_3,x}^{chiral})+{\bf V}\right),\begin{pmatrix}
{\bf G}_{5,7} & 0\\
0 & {\bf G}_{5,7}
\end{pmatrix} \right]+\frac{1}{2}\left(J(\mathrm{i}{\bf H}_{\omega_3,x}^{chiral})+{\bf V}\right)=
\begin{pmatrix}
{\bf G}_{1,7}+{\bf G}_{1,5} & 0\\
0 & 0
\end{pmatrix}+\begin{pmatrix}
{\bf G}_{1,5}-{\bf G}_{1,7} & 0\\
0 & 0
\end{pmatrix}\propto \begin{pmatrix}
{\bf G}_{1,5} & 0\\
0 & 0
\end{pmatrix},
\end{equation}
\end{widetext}
which is a basis element for the Lie algebra acting on the first enantiomer only.
Replacing $J(\mathrm{i}{\bf H}_{\omega_3,x}^{chiral})+{\bf V}$ with $J(\mathrm{i}{\bf H}_{\omega_3,x}^{chiral})-{\bf V}$ in \eqref{sum1} and \eqref{sum2}, we obtain a matrix proportional to $ \begin{pmatrix}
0 & 0\\
0 & {\bf G}_{1,5}
\end{pmatrix},$
which is a basis element for the Lie algebra acting on the second enantiomer only.	 
To conclude the proof, it suffices to compute commutators between these elements and the elements of Eq.~\eqref{firststep}, e.g.
$$ \left[\begin{pmatrix}
{\bf G}_{1,5} & 0\\
0 & 0
\end{pmatrix},\begin{pmatrix}
{\bf G}_{5,k} & 0\\
0 & {\bf G}_{5,k}
\end{pmatrix} \right]=\begin{pmatrix}
{\bf G}_{1,k} & 0\\
0 & 0
\end{pmatrix},$$
and 
$$ \left[\begin{pmatrix}
0 & 0\\
0 & {\bf G}_{1,5}
\end{pmatrix},\begin{pmatrix}
{\bf G}_{5,k} & 0\\
0 & {\bf G}_{5,k}
\end{pmatrix} \right]=\begin{pmatrix}
0 & 0\\
0 & {\bf G}_{1,k}
\end{pmatrix} $$
for all $k=1,\dots,7$. Since from the elements ${\bf G}_{1,k}$, we  obtain all
${\bf G}_{j,k}$, ${\bf F}_{j,k}$, and ${\bf D}_{j,k}$ using the relations \eqref{relation1}, \eqref{relation3}, and \eqref{relation4}, the Lie algebra generated by the five fields contained in $\mathcal{X}$ is  $\mathfrak{su}(7)\oplus\mathfrak{su}(7)$
which proves enantio-selective controllability. 

Summarizing, we have demonstrated for the $J=0/1/1$ subsystem that a single enantiomer is controllable with four fields with frequencies $\omega_1$ and $\omega_2$, while for enantio-selective control five fields containing the frequencies $\omega_1$, 
$\omega_2$, and $\omega_3=\omega_1+\omega_2$ are necessary. Here, we chose the polarizations to be $p_1=x$, $p_2=z$, $p_3=y$, $p_4=z$, and $p_5=x$. Other choices of the polarization directions also result in controllability and enantio-selective controllability as long as the pairs $p_1,p_2$ and $p_3,p_4$ are not the same and all three polarization directions $x$, $y$, $z$ are present. We obtain the same results for all  $J/J+1/J+1$-subsystems with $J\leq2$. The proof in the case $J=2/3/3$ is shown in the Appendix.
We show how to exploit these minimal sets of fields for the example of the carvone molecule in Sec.~\ref{subsec:carvoneJ1}.

\subsection{Controllability and enantio-selective 
excitation}
\label{subsec:subsets}

In view of the aim to achieve complete separation of two enantiomers in a mixture, complete enantio-selective controllability within the considered subsystem is a sufficient, but not a necessary condition. In the following, we construct an example where enantio-selective controllability within the set of  states that is reachable for a single enatiomer is sufficient to achieve enantio-selective excitation, that is, complete separation of the enantiomers. 

\begin{figure}
	\includegraphics[width=0.8\linewidth]{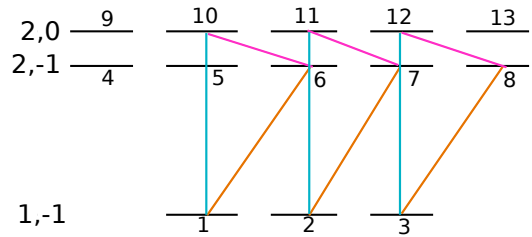} 
	\caption{Rotational subsystem containing the asymmetric top states $|1,\minusone,M\rangle$, $|2,\minusone,M\rangle$ and $|2,0,M\rangle$. The orange, pink, and turquoise lines indicate the transitions which are induced by fields with polarization $\sigma_+$ ($\sigma_-$, $z$)
	and frequencies $\omega_1$ ($\omega_2$, $\omega_3$), respectively.}
	\label{scheme_circular}
\end{figure}Consider a set of five control fields that generate the interaction Hamiltonians
\begin{equation*}
\{\mathrm{i}{\bf H}_{\omega_1,x},\mathrm{i}{\bf H}_{\omega_1,y},\mathrm{i}{\bf H}_{\omega_2,x},\mathrm{i}{\bf H}_{\omega_2,y},\mathrm{i}{\bf H}_{\omega_3,z}\},
\end{equation*}
with frequencies $\omega_1=E_{J+1,\tau'}-E_{J,\tau}$, $\omega_2=E_{J+1,\tau''}-E_{J+1,\tau'}$, and $\omega_3=E_{J+1,\tau''}-E_{J,\tau}$.
According to Sec.~\ref{subsec:subsystems},
a $J/J+1/J+1$-subsystem is not controllable with this set of fields since $(p_1=x,p_2=y) = (p_3=x,p_4=y)$. Instead of linearly polarized fields, we now consider left and right circularly polarized fields with polarization directions $\sigma_\pm = x \pm iy$. Assuming that the polarization directions of the fields with frequencies $\omega_1$ and $\omega_2$ are $\sigma_+$ and $\sigma_-$, respectively, the resulting anti-Hermitian interaction Hamiltonians are
\begin{equation*}
 \mathrm{i} {\bf H}_{\omega_1,\sigma_+} =  \mathrm{i} {\bf H}_{\omega_1,x} +  J ( \mathrm{i} {\bf H}_{\omega_1,y} )
\end{equation*}
and
\begin{equation*}
 \mathrm{i} {\bf H}_{\omega_2,\sigma_-} = \mathrm{i} {\bf H}_{\omega_2,x} - J(\mathrm{i} {\bf H}_{\omega_2,y})\,,
\end{equation*}
with 
\begin{equation}
J \left( \mathrm{i} {\bf H}_{\omega_i,a} \right) = \left[ \mathrm{i} {\bf H}_{0}, \mathrm{i} {\bf H}_{\omega_i,a}\right]/\omega_i.
 \label{J_H}
\end{equation}
The transitions induced by the set of interactions
\begin{equation*}
\{\mathrm{i}{\bf H}_{\omega_1,\sigma_+},\mathrm{i}{\bf H}_{\omega_2,\sigma_-},\mathrm{i}{\bf H}_{\omega_3,z}\},
\end{equation*}
are sketched in Fig.~\ref{scheme_circular} for the example of the subsystem~(i) of Fig.~\ref{spectrum_asym_top}. 
Obviously, the subsystem is not controllable by this set of fields, since some of the rotational states are not addressed at all. However, assuming that only the lowest manifold is populated initially, for the example of Fig.~\ref{scheme_circular} 
\begin{equation}
	\rho^{(\pm)}(0) = \frac{1}{3} \sum_{n=1}^3 |n\rangle \langle n |\,, 
	\label{initial_circ}
\end{equation}
enantio-selective excitation can be obtained by considering only
the set of reachable states, i.e., the set of rotational states to which population can be transferred by the control fields. These are the states labeled by $1$--$3$, $6$--$8$, and $10$--$12$ in Fig.~\ref{scheme_circular}. In this example, the set of reachable states is divided into three isolated subsystems, each consisting of three states. As a whole, the restriction of the system to such a set of reachable states is therefore not controllable either. 
However, a sufficient condition for enantio-selective excitation is that the three isolated subsystems are simultaneously enantio-selective controllable.
This requires the Lie algebra to be
\begin{equation}
\mathfrak{su}(3) \oplus \mathfrak{su}(3) \oplus  \mathfrak{su}(3) \oplus \mathfrak{su}(3) \oplus \mathfrak{su}(3) \oplus \mathfrak{su}(3)
\end{equation}
since each of the three-level systems is controllable if its Lie algebra is $\mathfrak{su}(3)$ and enantio-selective controllable if its Lie algebra is $\mathfrak{su}(3) \oplus \mathfrak{su}(3)$.

In order to determine the Lie algebra for a single enantiomer, we first consider the interaction Hamiltonians
\begin{eqnarray*}
\mathrm{i}{\bf H}_{\omega_1,\sigma_+}&\propto&\mu_a\Big({\bf G}_{1,6}+\sqrt{3}{\bf G}_{2,7}+\sqrt{6}{\bf G}_{3,8}\Big),\\ 
\mathrm{i}{\bf H}_{\omega_2,\sigma_-}&\propto&\mu_b\Big({\bf G}_{5,9}+{\bf G}_{8,12}+\sqrt{\frac32}\Big({\bf G}_{6,10}+{\bf G}_{7,11}\Big)\Big)  
\end{eqnarray*}
and show that, together with $\mathrm{i}{\bf H}_{0}$, they generate $\mathfrak{su}(3) \oplus \mathfrak{su}(3)  \oplus \mathfrak{su}(3)$.
Using Eq.~(\ref{relation4}), we find 
\[
J(\mathrm{i}{\bf H}_{\omega_1,\sigma_+}) \propto {\bf F}_{1,6}+\sqrt{3}{\bf F}_{2,7}+\sqrt{6}{\bf F}_{3,8}\,,
\]
with $J(\mathrm{i}{\bf H}_{\omega_1,\sigma_+})$  defined in Eq.~(\ref{J_H}).
Moreover, abbreviating commutators as 
$\mathrm{ad}_A B=[A,B]$, $\mathrm{ad}^{n+1}_A B=[A,\mathrm{ad}^{n}_A B]$, and $\mathrm{ad}^{0}_A B=B$, 
we note that 
\[
\mathrm{ad}^{2s}_J(\mathrm{i}{\bf H}_{\omega_1,\sigma_+})\mathrm{i}{\bf H}_{\omega_1,\sigma_+}\propto
{\bf G}_{1,6}+\sqrt{3}^{2s+1}{\bf G}_{2,7}+\sqrt{6}^{2s+1}{\bf G}_{3,8}
\]
with $s=0,1,2,\ldots$ Thus,
$$\begin{pmatrix}
\mathrm{ad}^{0}_{J(\mathrm{i}{\bf H}_{\omega_1,\sigma_+})} \mathrm{i}{\bf H}_{\omega_1,\sigma_+} \\
\mathrm{ad}^{2}_{J(\mathrm{i}{\bf H}_{\omega_1,\sigma_+})} \mathrm{i}{\bf H}_{\omega_1,\sigma_+} \\
\mathrm{ad}^{4}_{J(\mathrm{i}{\bf H}_{\omega_1,\sigma_+})} \mathrm{i}{\bf H}_{\omega_1,\sigma_+}
\end{pmatrix}=V \begin{pmatrix}
{\bf G}_{1,6}\\
{\bf G}_{2,7}\\
{\bf G}_{3,8}
\end{pmatrix}
$$
with
$$
 V=\begin{pmatrix} 
1 & \sqrt{3} &\sqrt{6} \\
1 & \sqrt{3}^{3} &\sqrt{6}^{3}\\
1 & \sqrt{3}^{5} &\sqrt{6}^{5} 
\end{pmatrix}. $$
The matrix $V$ is invertible since the entries $1,\sqrt{3},\sqrt{6}$ are all different which implies that
$
{\bf G}_{1,6}, {\bf G}_{2,7}, {\bf G}_{3,8} \in \mathrm{Lie}\{\mathrm{i}{\bf H}_{0} ,\mathrm{i}{\bf H}_{\omega_1,\sigma_+}\}.
$
From the commutation rules of the generalized Pauli matrices~\eqref{eq:commutators}, 
it follows that also
$$
{\bf X}_{1,6}, {\bf X}_{2,7}, {\bf X}_{3,8} \in \mathrm{Lie}\{\mathrm{i}{\bf H}_{0} ,\mathrm{i}{\bf H}_{\omega_1,\sigma_+}\}, \; {\bf X}\in\{{\bf G},{\bf F},{\bf D} \}.
$$
We then calculate the commutators
\[
\begin{split}
&[[\mathrm{i}{\bf H}_{\omega_2,\sigma_-},{\bf G}_{1,6}],{\bf G}_{1,6}]\propto {\bf G}_{6,10}, \\&
[[\mathrm{i}{\bf H}_{\omega_2,\sigma_-},{\bf G}_{2,7}],{\bf G}_{2,7}]\propto {\bf G}_{7,11}, \\&
[[\mathrm{i}{\bf H}_{\omega_2,\sigma_-},{\bf G}_{3,8}],{\bf G}_{3,8}]\propto {\bf G}_{8,12},
\end{split}
\]
and, using again the commutation relations of the generalized Pauli matrices and the rotational Hamiltonian, we find 
\[
\begin{split}
&{\bf X}_{6,10}, {\bf X}_{7,11}, {\bf X}_{8,12} \in \mathrm{Lie}\{\mathrm{i}{\bf H}_{0} ,\mathrm{i}{\bf H}_{\omega_1,\sigma_+},\mathrm{i}{\bf H}_{\omega_2,\sigma_-}\}, \\&{\bf X}\in\{{\bf G},{\bf F},{\bf D} \}.
\end{split}
\]
Since 
\[
\begin{split}
&\mathrm{Lie}\Big\{{\bf X}_{1,6}, {\bf X}_{2,7}, {\bf X}_{3,8},{\bf X}_{6,10}, {\bf X}_{7,11}, {\bf X}_{8,12}\mid {\bf X}\in\{{\bf G},{\bf F},{\bf D} \} \Big\} \\&
\cong \mathfrak{su}(3)\oplus  \mathfrak{su}(3)\oplus  \mathfrak{su}(3),
\end{split}
\]
we have proven that the three isolated three-level systems are simultaneously controllable with the interaction operators $\mathrm{i}{\bf H}_{\omega_1,\sigma_+}$ and $\mathrm{i}{\bf H}_{\omega_2,\sigma_-}$.

To obtain enantio-selective control of each of these three cycles, we consider the interaction with the third field, namely
\[
\begin{split}
\mathrm{i}{\bf H}_{\omega_3,z}\propto\mu_c\Big({\bf G}_{2,11}+\frac{\sqrt{3}}{2}\Big({\bf G}_{3,12}+{\bf G}_{1,10}\Big)\Big)\,,
\end{split}
\]
or, for the composite system consisting of the two enantiomers,
$$
\mathrm{i}{\bf H}_{\omega_3,z}^{chiral}=(\mathrm{i}{\bf H}_{\omega_3,z})\oplus(-\mathrm{i}{\bf H}_{\omega_3,z}).
$$
The interaction operators 
$$
\mathrm{i}{\bf H}_{\omega_i,a}^{chiral}=(\mathrm{i}{\bf H}_{\omega_i,a})\oplus(\mathrm{i}{\bf H}_{\omega_i,a})
$$
for $(\omega_i,a)=(\omega_1,\sigma_+)$ and $(\omega_2,\sigma_-)$ together with $\mathrm{i}{\bf H}_{0}^{chiral}$ create, among others, the operators ${\bf G}_{1,6}\oplus {\bf G}_{1,6}$ and ${\bf G}_{1,10}\oplus {\bf G}_{1,10}$. We compute the double bracket  
\[
[[\mathrm{i}{\bf H}_{\omega_3,z}^{chiral},{\bf G}_{1,6}\oplus {\bf G}_{1,6}],{\bf G}_{1,6}\oplus {\bf G}_{1,6}]\propto {\bf G}_{1,10}\oplus (-{\bf G}_{1,10}),
\]
and taking the sum and difference with ${\bf G}_{1,10}\oplus {\bf G}_{1,10}$, the operators ${\bf G}_{1,10}\oplus {\bf 0}$ and ${\bf 0}\oplus {\bf G}_{1,10}$ are generated. In the same manner, all operators
\begin{equation}
{\bf X}_{i,j}\oplus {\bf 0} \,\, \mbox{and} \,\, {\bf 0}\oplus {\bf X}_{i,j} \,\, \mbox{for} \,\, {\bf X}\in\{{\bf G},{\bf F},{\bf D} \}
\label{Xij_chiral}
\end{equation} 
can be generated. Since the operators ${\bf X}_{i,j}$ form a basis of the Lie algebra $\mathfrak{su}(3)\oplus  \mathfrak{su}(3)\oplus  \mathfrak{su}(3)$,
the operators (\ref{Xij_chiral}) span
$\mathfrak{su}(3)\oplus  \mathfrak{su}(3)\oplus  \mathfrak{su}(3) \oplus \mathfrak{su}(3)\oplus  \mathfrak{su}(3)\oplus  \mathfrak{su}(3)$, and thus the three three-level systems are simultaneously enantio-selective controllable.

As a result, for the initial state (\ref{initial_circ}), complete enantio-selective excitation can be obtained by two circularly polarized and one linearly polarized fields. As we shall see for the example of carvone in Sec.~\ref{subsec:carvone:circ}, those isolated three-level systems are particularly suited for enantio-selective excitation in real molecules.

\section{Application: Complete enantiomer-specific state transfer in carvone}
\label{sec:carvone}

We now show how to use the mathematical results of Sec.~\ref{subsec:subsystems} and~\ref{subsec:subsets} to derive actual pulse sequences for enantiomer-specific state transfer in microwave three-wave mixing spectroscopy, using the example of carvone molecules. Our choice of example is motivated by experiments demonstrating 6\% enantiomeric enrichment for this molecule~\cite{PerezAngewandte17}, the largest enrichment obtained with three-wave mixing spectroscopy so far.
We simulate the rotational dynamics of $R$- and $S$-carvone by numerically solving the rotational Schr\"odinger equation~\eqref{quantumcontrol}~\footnote{ 
Note that our simulations of the coherent dynamics do not explictly take the rotational temperature into account.  Finite rotational temperature reduces the purity of the initial quantum state and thus the degree of selectivity that can be obtained by any coherent dynamics. Our pulse sequences will induce the maximal degree of enantio-selectivity that is compatible with the purity of the initial thermal ensemble, and temperature can thus simply be factored in.}
with rotational constants 
$A=2237.21\,$MHz,  $B=656.28\,$MHz, and $C=579.64\,$MHz, and dipole moments $\mu_a=2.0\,$D $\mu_b=3.0\,$D, and $\mu_c=0.5\,$D~\cite{MorenoStructural13}.

We present two different strategies, in Sec.~\ref{subsec:carvoneJ1} and~\ref{subsec:carvone:circ}, to achieve complete enantiomer-specific state transfer for population initially distributed over $M$-degenerate states. The working principle of both strategies is to combine enantio-selectivity (due to the sign difference in one of the dipole moments) with an energetic separation of population residing initially in degenerate states. In Sec.~\ref{subsec:carvoneJ1}, we use the insight  on the number of fields required for \textit{complete} controllability from Sec.~\ref{subsec:subsystems} and construct pulse sequences to energetically separate (i) $M$-degenerate states of one enantiomer and (ii) the two enantiomers (in $M$-degenerate states) in a racemic mixture. The pulses  drive transitions within the $J=0/J=1/J=1$ rotational submanifold, cf. Fig.~\ref{level_scheme_J0J1}. Even in this comparatively small manifold, the pulse sequence for enantiomer-selective population transfer consists of 12 pulses sampled from five different fields, i.e., five different combinations of polarization directions and frequencies. In order to obtain a simpler sequence, 
we forego complete controllability in Sec.~\ref{subsec:carvone:circ} and use the insight from Sec.~\ref{subsec:subsets} to achieve  enantiomer-selective excitation in the presence of $M$-degeneracy. 
A sequence of three pulses corresponds to partitioning the rotational submanifold into isolated subsystems and drives simultaneously several three-wave mixing cycles.
For this strategy to succeed, the initial rotational submanifold needs to have the smallest degeneracy factor $g_J=2J+1$. We therefore consider transitions within the $J=1$/$J=2$/$J=2$ rotational submanifold in Sec.~\ref{subsec:carvone:circ}.

\subsection{Control and enantiomer-selective control exploiting complete controllability}
\label{subsec:carvoneJ1}
\begin{figure}
 \includegraphics[width=1.0\linewidth]{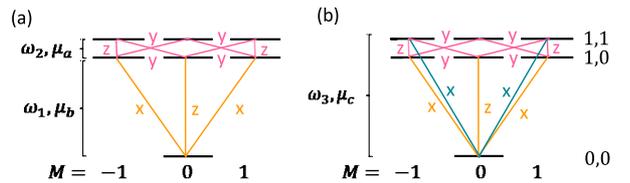} 
  \caption{Level scheme of the rotational subsystem consisting of the states $|0,0,0\rangle$, $|1,0,M\rangle$, and $|1,1,M\rangle$ with $M=\minusone,0,1$. The orange and pink lines in panel (a) indicate the four fields which yield complete controllability of this subsystem for a single enantiomer. The polarization of the fields is denoted by 
$x$, $y$, and $z$. The additional field which is required for enantio-selective control is indicated in panel (b) by turquoise lines.}
	\label{level_scheme_J0J1}
\end{figure}

The simplest rotational subsystem that allows for enantiomer-selective population transfer using three-wave mixing spectroscopy 
consists of the rotational states $|J,\tau, M \rangle = |0,0,0 \rangle$, $|1,0,M \rangle$, and $|1,1,M \rangle$ with
$M=\minusone,0,1$, and rotational energies $E_{J,\tau}=E_{00}$, $E_{10}$, and $E_{11}$, cf. Fig.~\ref{level_scheme_J0J1}. A single enantiomer is completely controllable with four (microwave) fields, as we have shown in Sec.~\ref{subsec:subsystems}, for example with two fields with frequency $\omega_1= (E_{10}-E_{00})/\hbar$ and $x$-, respectively  $z$-polarization and two fields with frequency $\omega_2= (E_{11}-E_{10})/\hbar$ and  $y$-, respectively $z$-polarization. The transitions induced by these fields are indicated by orange and pink lines in Fig.~\ref{level_scheme_J0J1}(a); they form closed loops connecting the four states $|0,0,0 \rangle$, $|1,0,\pm 1 \rangle$, $|1,1,\pm 1 \rangle$, and $|1,0,0 \rangle$. 
Complete controllability implies that population in any initial state 
within the rotational manifold can be driven into any other initial state within that manifold. This means in particular that population in degenerate states, for example $|1,0,\pm 1 \rangle$,  can be driven into states with different energy. Such an energetic separation can serve as precursor for complete enantio-selective excitation, as we show below.
It also has further applications and could, for example, be used towards purifying an incoherent ensemble with electric fields only or distilling a specific molecular orientation.

\begin{figure*}
  \includegraphics[width=1.0\linewidth]{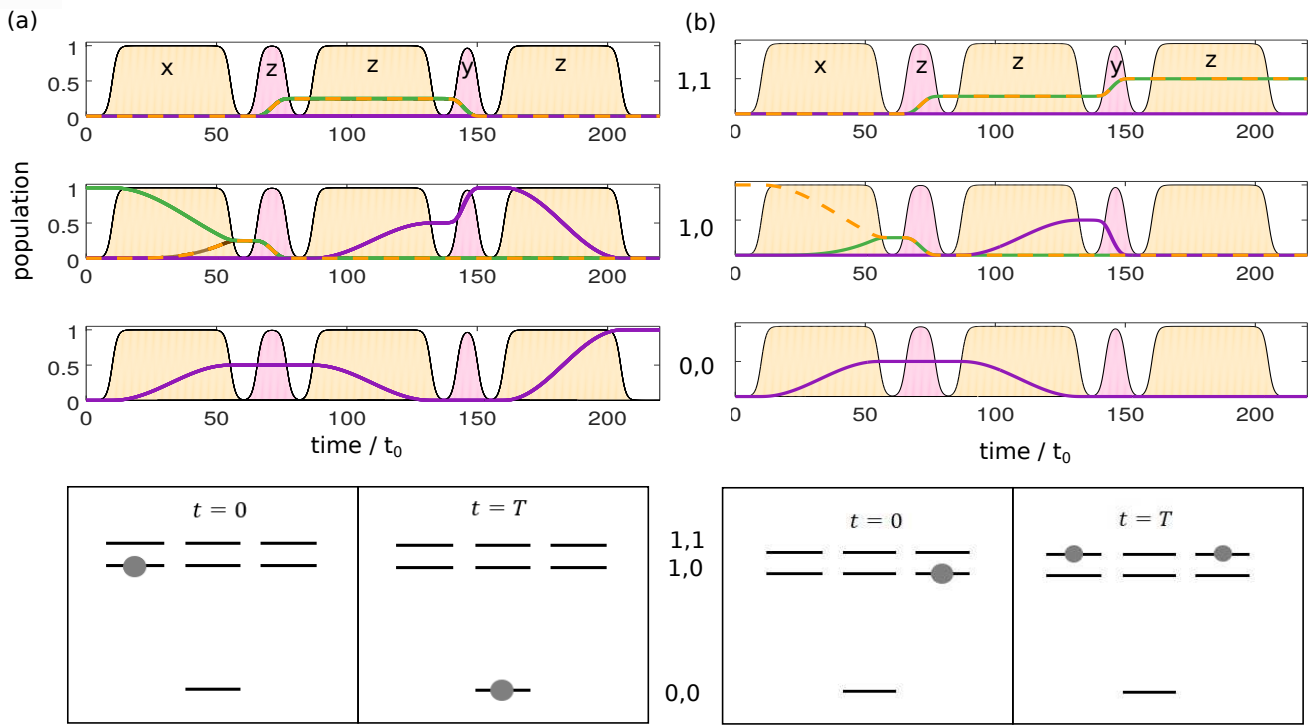} 
  \caption{Rotational dynamics energetically separating the degenerate initial states $|1,0,\minusone\rangle$ (a) and$|1,0,1\rangle$ (b) for a single enantiomer. Top: Population of the rotation states $|0,0,0\rangle$ (lowest panels), $|1,0,M\rangle$ (middle panels), and $|1,1,M\rangle$ (upper panels). The population dynamics of the degenerate states are depicted by green ($M=\minusone$), purple ($M=0$), and orange ($M=1$) lines. The envelope of the pulses is indicated by the orange ($\omega=\omega_1$) and pink ($\omega=\omega_2$) shapes, $x$, $y$, and $z$ indicates the polarization of the corresponding fields. Time is given in units of $t_0=\hbar/B$. Bottom: Sketch of the initial ($t=0$) and final ($t=T$) states, marked by gray dots.}
	\label{sim_norm_control}
\end{figure*}
We therefore consider the following control problem for a single enantiomer: Given that the initial state is an incoherent ensemble of the two degenerate  $|1,0,M \rangle$ states, 
\begin{equation}
\rho(0) = \frac{1}{2} |1,0,\minusone \rangle \langle 1,0,\minusone | + \frac{1}{2} |1,0,1 \rangle \langle 1,0,1 | \,,
\end{equation}
find a pulse sequence that drives the population with $M=+1$ into a final state with different rotational energy than the $M=\minusone$ component. 
As an example, we have chosen $|0,0,0\rangle$ and  $1/\sqrt{2} (|1,1,\minusone\rangle + |1,1,1\rangle)$ as target states.
The initial and desired final states are sketched as gray dots in the bottom panels of Fig.~\ref{sim_norm_control}, the 
upper panel of which shows 
the pulse sequence that drives the corresponding rotational dynamics. 
In detail, starting from the initial states  $|1,0,\minusone \rangle$ (see Fig.~\ref{sim_norm_control}(a)) and $|1,0,1 \rangle$ (see Fig.~\ref{sim_norm_control}(b)), the state $|1,0,0 \rangle$ (purple line in the middle panel) can be reached by two different excitation pathways: {\it via} the states $|1,1,\pm 1 \rangle$ and {\it via} $|0,0,0 \rangle$. The 1st, 2nd, and 4th pulse transfer 50\% of the population to state $|1,0,0\rangle$ {\it via} the first pathway, while pulses 1 and 3 transfer the other half of the initial population along the second pathway. Interference between the two pathways in $|1,0,0\rangle$ is constructive for the initial state $|1,0,\minusone\rangle$ and destructive for the initial state  $|1,0,1\rangle$ (see purple lines in the middle panel of Figs.~\ref{sim_norm_control}(a) and (b) near $t=150\,t_0$). Therefore, the initial state $|1,0,\minusone\rangle$ is transferred to $|1,0,0\rangle$ while the initial state $|1,0,1\rangle$   is transferred to $1/\sqrt{2} (|1,1,\minusone\rangle + |1,1,1\rangle)$ at the end of pulse 4. Finally, the 5th pulse transfers the population from $|1,0,0\rangle$ to the desired final state $|0,0,0\rangle$ in Fig.~\ref{sim_norm_control}(a) while not affecting the population in $|1,1,\pm 1\rangle$
Fig.~\ref{sim_norm_control}(b).
The two initially degenerate states thus become energetically separated using four fields, with two different frequencies and two polarization components.

\begin{figure*}
  \includegraphics[width=1.0\linewidth]{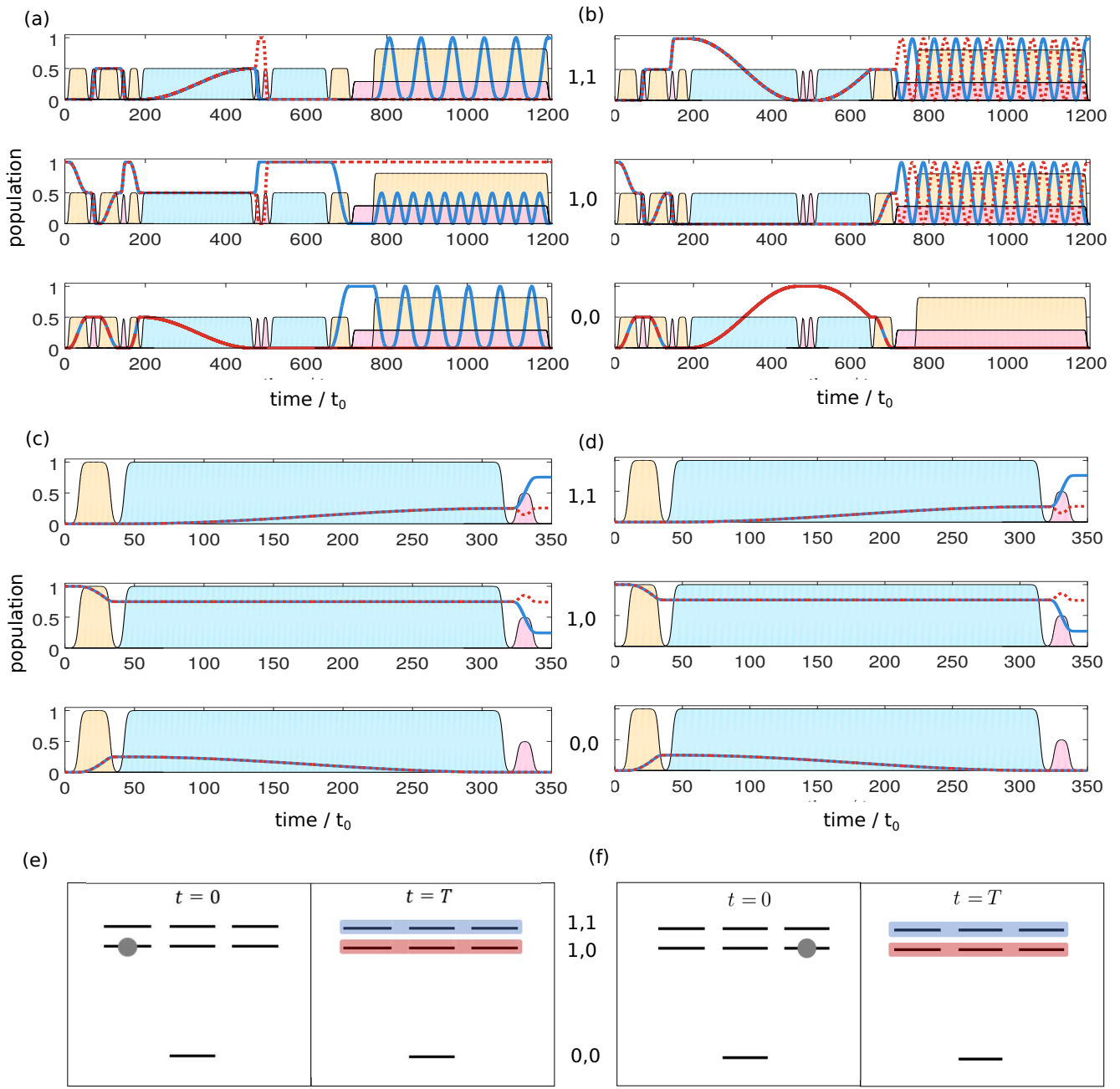} 
  \caption{Full control of enantiomer-selective state transfer despite $M$-degeneracy using five different fields (a,b). The complete pulse sequence with the five fields in (a,b) reads 1. ${\bf H}_{\omega_1,x}$, 2. ${\bf H}_{\omega_2,z}$, 3. ${\bf H}_{\omega_1,z}$, 4. ${\bf H}_{\omega_2,y}$, 5. ${\bf H}_{\omega_1,z}$, 6. ${\bf H}_{\omega_3,x}$, 7. ${\bf H}_{\omega_2,y}$, 8. ${\bf H}_{\omega_2,z}$, 9. ${\bf H}_{\omega_3,x}$, 10. ${\bf H}_{\omega_1,z}$, 11. ${\bf H}_{\omega_2,y}$, and 12. ${\bf H}_{\omega_1,x}$.
  	The two enantiomers are denoted by solid blue and dashed red lines. For comparison: Incomplete enantiomer-selective state transfer due to $M$-degeneracy in standard three-wave mixing cycles, using the fields ${\bf H}_{\omega_1,x}$, ${\bf H}_{\omega_2,z}$, and ${\bf H}_{\omega_3,y}$ (c,d). 
  	Panels (a,c) depict the rotational dynamics for the initial state $|1,0,\minusone\rangle$ and (b), (d) those for the initial state $|1,0,1\rangle$ with the sup-panels showing the accumulated population of the rotational states $|0,0,0\rangle$ (lowest panels), $|1,0,M\rangle$ (middle panels), and $|1,1,M\rangle$ (top panels). The pulse envelopes are indicated by orange ($\omega=\omega_1$), pink ($\omega=\omega_2$), and turquoise ($\omega=\omega_3$) shapes. The height of these shapes indicates the maximal electric field strength (in arbitrary units). Time is given in units of $t_0=\hbar/B$.
Panels (e) and (f) illustrate the initial ($t=0$) and final ($t=T$) populations with gray indicating both enantiomers in the same state  and blue and red representing the two (separated) enantiomers.
}
   \label{sim_enantio_selective_control}
\end{figure*}

For enantiomer-selective control, an additional field with frequency $\omega_3=\omega_1+\omega_2$ is required to allow for three-wave mixing. In our example, we choose $x$-polarization for $\omega_3$ such that we have 
three mutually orthogonal fields with 
${\bf H}_{\omega_1,z}$ (central orange line in Fig.~\ref{level_scheme_J0J1} (b)), ${\bf H}_{\omega_2,y}$ (pink lines), and ${\bf H}_{\omega_3,x}$ (turquoise lines). 
If the initial state is the ground rotational state, three-wave mixing results in complete separation of the enantiomers into energetically separated levels~\cite{Leibscher19}. This requires, however, preparation of the molecules close to zero temperature. For typical experimental conditions, the initial state has to be chosen with $J>0$~\cite{EibenbergerPRL17,PerezAngewandte17} 
and thus contains degenerate rotational states. Then, three fields are not sufficient to obtain complete enantio-selectivity. 
Therefore, we  consider the initial ensemble~\eqref{eq:rho-racemic}
 with
 \begin{equation}
 \rho^{(\pm)}(0)= \frac{1}{2} \left( |1,0,\minusone \rangle \langle 1,0,\minusone | + |1,0,1 \rangle \langle 1,0,1 | \right).
    \label{density0_pm} 
 \end{equation}  
 The initial states $|1,0,\minusone \rangle$ and $|1,0,1 \rangle$ are depicted in Fig.~\ref{sim_enantio_selective_control}(e) and (f) with the gray circles indicating that both enantiomers occupy the same states. The control aim is to drive the two enantiomers into rotational states with different energies, cf. the red and blue shades in Fig.~\ref{sim_enantio_selective_control}(e) and (f).
 
The combination of fields ${\bf H}_{\omega_1,z}$, ${\bf H}_{\omega_2,y}$, and ${\bf H}_{\omega_3,x}$, indicated in Fig.~\ref{level_scheme_J0J1}(b), which works if the initial state is $\ket{0,0,0}$, obviously fails for Eq.~\eqref{eq:rho-racemic} since it does not create three-wave mixing cycles for the $\ket{1,0,M}$ states.
This can be remedied by choosing instead a sequence containing the fields ${\bf H}_{\omega_1,x}$, ${\bf H}_{\omega_2,z}$, and ${\bf H}_{\omega_3,y}$. However, due to insufficient controllability with three fields in the presence of $M$-degeneracy, the population transfer is only partially enantiomer-selective, cf. the corresponding rotational dynamics in Fig.~\ref{sim_enantio_selective_control}(c) and (d), where the solid blue and dashed red lines present the two enantiomers. For complete
enantio-selective excitation, all five fields depicted in  Fig.~\ref{level_scheme_J0J1}(b) are required, as is illustrated by Fig.~\ref{sim_enantio_selective_control}(a) and (b).

The pulse sequence, which leads to complete separation of the enantiomers into energetically separated levels, consists of 12 pulses: The first four pulses are the same as the pulse sequence shown in Fig.~\ref{sim_norm_control}. Transferring the initial states $|1,0,\minusone\rangle$ and $|1,0,1\rangle$ into $|1,0,0\rangle$, respectively $1/\sqrt{2} (|1,1,\minusone\rangle + |1,1,1\rangle)$, they lead to an energetic separation of the two initially degenerate $M$ states,  but are not yet enantiomer-selective. Two more pulse sequences realize enantiomer-selective three-wave mixing cycles for the two initial states separately. First, enantiomer-selective transfer for the initial state $|1,0,\minusone\rangle$ is obtained by three-wave mixing with the fields ${\bf H}_{\omega_1,z}$,  ${\bf H}_{\omega_3,x}$, and ${\bf H}_{\omega_2,y}$ (pulses 5, 6, and 7). Analogously, pulses 9, 10, and 11 form a three-wave mixing cycle
for the initial state $|1,0,1\rangle$. After pulse 11 the enantiomers of both initial states are separated in energy. The two cycles for the different $M$-states are synchronized by applying pulse 12 (in addition to pulse 11), such that all population of one enantiomer  is collected in the highest rotational state (blue lines) while all population of the other enantiomer  is excited to the intermediate level (dashed red lines). 
Fig.\ref{sim_enantio_selective_control}(a) and (b) thus confirms complete enantio-selective state transfer in a racemic mixture of initially degenerate $M$-states for a set of microwave fields for which enantio-selective controllability is predicted in Sec.~\ref{subsec:subsystems}.  

The analysis of enantio-selective controllability of Sec.~\ref{sec:controllabilityresults} B yields the minimal number of different fields which are required for enantiomer-selective population transfer, but does not make any predictions about the temporal shape of the fields. In particular, it does not predict the number of individual pulses. The control sequence shown in Fig. \ref{sim_enantio_selective_control}(a) and (b) contains 12 individual pulses applied either sequentially or partially overlapping. 
Here, complete enantio-selectivity is obtained by constructing an individual three-wave mixing cycle for every initial state. This implies that population  initially in the degenerate $M$-states first has to be separated in energy so that they can be addressed individually. If the degeneracies become larger (for higher $J$), the pulse sequences become more complicated, because more degenerate states have to be separated in energy and three-wave mixing cycles for each of these states have to be constructed.  
Such pulse sequences may experimentally not be feasible or at least technically very challenging to implement. This is true in particular for rotational
subsystems with higher rotational quantum numbers as in earlier microwave three-wave mixing experiments ~\cite{PerezAngewandte17}, where cycles with $J=1/2/2$ or $J=2/3/3$ have been addressed because of their better frequency match and higher Boltzmann factors. For these cases, a control strategy based on partitioning the rotational manifold into subsystems, as discussed in Sec.~\ref{subsec:subsets}, may be better suited. This will be discussed next.

\subsection{Complete enantiomer-selective population transfer using synchronized three-wave mixing}
\label{subsec:carvone:circ}

\begin{figure*}[tb]
  \includegraphics[width=\linewidth]{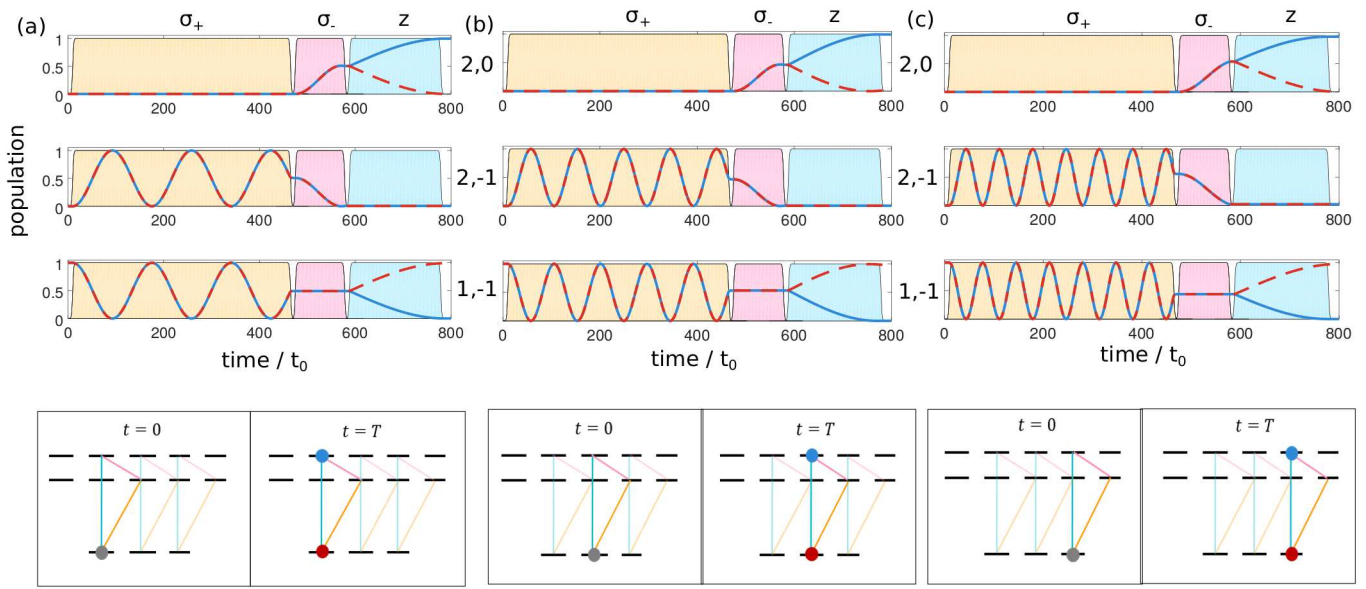} 
  \caption{Full control of enantiomer-selective state transfer, based on synchronized three-wave mixing with  ${\bf H}_{\omega_1,\sigma_+}$, ${\bf H}_{\omega_2,\sigma_-}$, ${\bf H}_{\omega_3,z}$ ($z-$polarization). Panels (a), (b), and (c) depict the rotational dynamics for the initial states  with $M=\minusone$, $M=0$, and $M=1$, respectively, with the overall
population in $|1\,\minusone\,M\rangle$, $|2\,\minusone\, M\rangle$, and
$|2\, 0\, M\rangle$ shown in the upper, middle, and lower sub-panels.
The two enantiomers are denoted by solid blue and dashed red lines. The envelope of the pulses is indicated by the orange ($\omega=\omega_1$), pink ($\omega=\omega_2$), and turquoise ($\omega=\omega_3$) shapes. Time is given in units of $t_0=\hbar/B$. Bottom: Sketch of the initial ($t=0$) and final ($t=T$) states with  gray circles indicating  both enantiomers in the same state, blue and red circles representing  the two (separated) enantiomers. The transitions induced by the three fields are indicated by the orange, pink, and turquoise lines. The transitions that affect the initial states with $M=\minusone$ (a), $M=0$ (b), and $M=1$ (c) are highlighted.}
	\label{sim_control_J1J2}
\end{figure*}
Another route to enantiomer-selective state transfer is provided by partitioning the relevant rotational manifold into subsystems that form
individual three-wave mixing cycles and uncontrollable ``satellites'', as discussed in Sec.~\ref{subsec:subsets}. Provided that the initial state contains population only within the various three-level cycles, the lack of complete controllability does not preclude enantiomer-selective population transfer. In other words, one needs to consider manifolds $\ket{J,\tau,M_J}$, $\ket{J^\prime, \tau', M_{J^\prime}}$, $\ket{J^\prime, \tau^\prime, M_{J^\prime}}$ where $J<J^\prime$ and choose the transitions realizing the three-wave mixing such that the initial state resides in the manifold with lower $J$.
An advantage of this approach is that three different fields, if properly chosen, are sufficient. 

As an experimentally relevant example, we consider the rotational subsystem made up of $\{\ket{1,\minusone,M}\}$, $\{|2,\minusone,M\rangle\}$, and $\{|2,0,M\rangle\}$ and construct a pulse sequence that achieves complete enantiomer-selective population transfer despite $M$-degeneracy. We assume that, initially, only the lowest rotational levels, those with $J=1$, are populated. The racemic mixture is then described by Eq.~\eqref{eq:rho-racemic} with 
\begin{eqnarray}
  \rho^{(\pm)}(0)= \frac{1}{3} \big(&& |1, \minusone, \minusone \rangle \langle 1, \minusone, \minusone | \\
                                    &&+ |1, \minusone, 0 \rangle \langle 1, \minusone, 0 | + 
|1, \minusone, 1\rangle \langle 1, \minusone, 1| \big)\,.  \nonumber
\end{eqnarray} 
Applying a standard three-wave mixing pulse sequence with linearly polarized fields with orthogonal polarization directions results at most in about 80\% enantio-selectivity (data not shown). In contrast, the circularly polarized fields discussed in  Sec.~\ref{subsec:subsystems} allow for a complete separation of the enantiomers. This can be seen in Fig.~\ref{sim_control_J1J2}. 

The three subsystems which are isolated by applying left- and right-circularly polarized light are indicated in the bottom panels of Fig.~\ref{sim_control_J1J2}: The field with $\sigma_+$-polarization (orange line) induces transitions between 
$|1,\minusone, M \rangle$ and $|2,\minusone, {M\!+\!1}\rangle$, while the $\sigma_-$-polarized field (pink line) drives transitions between $|2,\minusone,M \rangle$ and $|2,0,M\!\minusone \rangle$, and the linearly $z$-polarized field (turquoise line) closes the cycles.
For all the initially populated, degenerate $M$-states, the population is thus  trapped into a three-level subsystem and cannot spread over the whole manifold, as it would happen when using three linearly polarized fields with orthogonal polarization directions. 

The corresponding rotational dynamics is depicted in the upper panels of Fig.~\ref{sim_control_J1J2}(a)--(c). The pulse sequence that leads to complete enantio-selective excitation is essentially a three-wave mixing cycle: The first pulse creates a 50/50 coherence between the ground and first excited rotational level of each three-level system.  The second pulse transfers the population from the intermediate state to the highest state and the third, $z$-polarized pulse induces the enantiomer-specific interference between the ground state and highest excited state. There is, however, an important difference to the standard three-wave mixing cycles used so far --- the pulses are chosen such that they synchronize the three subsystems, allowing to reach a 50/50 coherence between the ground and first excited  state for each of the subsystems. As can be seen in Fig.~\ref{sim_control_J1J2}, the Rabi angles of each subsystem are different,
due to the different Clebsch-Gordon coefficients, respectively the different elements of the Wigner $D$-matrix, in Eq.~\eqref{transition_asym}.
A 50/50 coherence for all three subsystems occurs after three Rabi oscillations for the subsystem depicted in (a), 5 oscillations for (b), and 7 oscillations for (c). The synchronized three-level cycles then lead to complete separation of the enantiomers into energetically separated levels, by applying a sequence of only three pulses, cf. Fig.~\ref{sim_control_J1J2}.

When choosing the pulse amplitude and duration, it is important to realize that the subsystems undergo either all an even or all an odd number of Rabi oscillations, so that they accumulate the same phase. Otherwise, the interference effects induced by the third pulse will cancel each other. 
This excitation scheme can be easily extended to rotational manifolds with larger $J$, since the manifolds can always be broken up into isolated subsystems where three pulses are sufficient to energetically separate  the enantiomers. The number of pulses is thus independent of the number of degenerate states in the initial ensemble. The pulse duration of the first pulse may have to be longer (or its amplitude larger), since, for larger $J$, this pulse needs to synchronize  Rabi oscillations of more three-level cycles. However, this does not pose a fundamental difficulty. Synchronized three-wave mixing cycles driven with two circularly  polarized and one linearly polarized field should thus enable complete enantiomer-selective population transfer in microwave three-wave mixing experiments.

\section{General design principles}\label{sec:strategies}

Figures~\ref{sim_norm_control}, \ref{sim_enantio_selective_control},  and~\ref{sim_control_J1J2}  show three pulse sequences achieving $M$-sensitive, respectively enantiomer-selective, population transfer.
Each of these sequences represents only one among many possible solutions to the respective control problem. One could, for example, replace our combination of $\pi$- and $\pi/2$-pulses by a sequence inducing adiabatic passage~\cite{KralPRL01,LiPRL07,LiJCP10} or by one derived from shortcuts to adiabaticity~\cite{Vitanov19}. When adapting a given pulse sequence designed to start from the non-degenerate $J=0$-level to addressing a degenerate one ($J>0$), the following design principles will ensure selectivity despite  $M$-degeneracy.

First, one needs to select the appropriate combination of frequencies and polarizations, as discussed in Sec.~\ref{sec:controllabilityresults}, i.e.,   four different fields including all three linear polarization directions and two resonant frequencies for complete rotational controllability in a $J/J+1/J+1$ manifold; five different fields including all three linear polarization directions and three resonant frequencies for complete enantio-selective controllability in a $J/J+1/J+1$ manifold; and three different fields with three resonant frequencies, two with opposite circular polarization directions and one linearly polarized one, for enantio-selective control in ``parallel'' three-level cycles. The specific choice of the fields determines the states that will be addressed.

The pulse sequence then needs to be chosen such that it creates closed cycles for population transfer and  constructive, respectively destructive,  interference. The case most similar to three-wave mixing starting from $J=0$ is enantiomer-selective population transfer in ``parallel'' three-level cycles, cf. Sec.~\ref{subsec:carvone:circ}, where the replacement of linear by circular polarization for two of the fields breaks the symmetry between transitions with $M\leftrightarrow M+1$ and those with $M\leftrightarrow M-1$. All that is required in addition is synchronization of the cycles due to the $M$-dependent transition matrix elements. The interference for enantio-selectivity is achieved as before~\cite{KralPRL01,LiPRL07,Leibscher19,HirotaPJA12}. 
In case of rotational state transfer with $M$-selectivity, Fig.~\ref{sim_norm_control} in Sec.~\ref{subsec:carvoneJ1}, four states (in three levels) are involved since four fields are required for complete controllability. The sequence is chosen such that it creates constructive and destructive interference for states with opposite $M$. In order to generalize our example in Fig.~\ref{sim_norm_control}, with initial population in the states $M=\pm 1$, to higher degeneracies, one would need to combine  $\pm M$-selectivity with synchronization, to account for the $|M|$-dependent transition matrix elements. Finally, pulse sequences,  based on complete controllability, driving  enantiomer-selective  population transfer in a mixture of degenerate rotational states concatenate $M$-selective four-state cycles with enantio-discriminating three-level cycles, as in Fig.~\ref{sim_enantio_selective_control}. 

\section{Conclusions}      \label{sec:concl}

We have used Lie-algebraic techniques of controllability analysis to determine the number and type (in terms of frequency and polarization direction) of electric fields that allow to completely control the rotational dynamics of an asymmetric top molecule, despite the degeneracy with respect to the orientational quantum number $M$. This result in itself is already remarkable --- it implies that it is not necessary to lift the degeneracy with e.g. a magnetic field in order to selectively address each rotational level. Rather, selectivity can be achieved by exploiting differences of the transition matrix elements, using four different combinations of frequency and polarization direction. To demonstrate how this type of controllability can be utilized, we have constructed a pulse sequence that energetically separates population incoherently distributed over degenerate levels, as a precursor for distilling a specific molecular orientation. Exploiting complete controllability of rotational states despite the $M$-degeneracy may also be helpful for laser cooling of asymmetric molecules~\cite{AugenbraunPRX20} or their use in robust qubit encodings~\cite{AlbertPRX20}.

We have then introduced the concept of enantio-selective controllability, in order to analyze simultaneous controllability of the two enantiomers of a chiral molecule, driven by the same set of external fields. 
This analysis was motivated by microwave three-wave mixing spectroscopy aiming to energetically separate enantiomers in a racemic mixture, with current protocols suffering from population loss due to partially incomplete three-level cycles~\cite{EibenbergerPRL17,PerezAngewandte17,PerezJPCL18}. 
We have proven that complete enantio-selective controllability can be achieved with five different, suitably chosen combinations of frequency and polarization direction. This result implies the existence of microwave three-wave mixing protocols that allow for complete enantiomer-selective population transfer despite the $M$-degeneracy.
It is also relevant for all other enantiomer-specific processes which rely on  rotational dynamics, such as the non-resonant excitation of rotational wave packets by interaction with induced dipole moments \cite{YachmenevPRL16,TutunnikovJPCL18,MilnerPRL2019,TutunnikovPRA2020}.

For the example of microwave three-wave mixing, knowledge of the relevant light-matter couplings has allowed us to design a pulse sequence which drives enantiomer-specific population transfer. Our numerical simulations of the rotational dynamics for the example of carvone confirm nearly 100\% enantio-selectivity. The sequence consists of 12 pulses, sampled from five fields driving the same type of transitions as those used in the earlier microwave  experiments~\cite{EibenbergerPRL17,PerezAngewandte17,PerezJPCL18}. 
Admittedly, the pulse sequence is rather complicated, even for the smallest rotational subsystem. Therefore, we have identified, based on the controllability analysis of subsets of states, an alternative control strategy that relies on isolating ``parallel'' three-level subsystems for each degenerate level in a single manifold.  We have shown with numerical simulations 
that simultaneous control of the isolated subsystems yields complete enantio-selective excitation with a much simpler protocol containing only three fields, chosen to synchronize the population transfer in all of the cycles. 
The corresponding pulse sequence requires one left-circularly, one right-circularly, and one linearly polarized field and is within the capabilities of current microwave technology. Our proposal thus eliminates an important obstacle toward complete enantiomer-selective state transfer in three-wave mixing experiments.

More broadly, our work testifies to the value of mathematical controllability analysis in general and the Lie--Galerkin approximation in particular for topical problems in quantum control. The same techniques can in principle also be applied to  many-body dynamics or open quantum systems, where the spectral gap condition required to invoke the Lie--Galerkin approximation will translate into a timescale separation argument.  It will be interesting to see in these cases how far controllability despite degeneracy can be pushed. While in our  example of asymmetric quantum rotors, the key to controllability despite degeneracy is found in the 3D nature of the light-matter coupling, it is presently an open question which mechanisms could be leveraged for the control of  many-body dynamics or open quantum systems.

\begin{acknowledgments}
  We gratefully acknowledge financial support from the Deutsche Forschungsgemeinschaft through CRC 1319 ELCH and from the
European Union's Horizon 2020 research and innovation programme
under the Marie Sklodowska-Curie grant agreement Nr. 765267 (QuSCo). 
MS and UB also thank the ANR projects SRGI ANR-15-CE40-0018 and Quaco ANR-17-CE40-0007-01.
\end{acknowledgments}
   
\appendix* 
\section{Enantio-selective controllability of $J=2/3/3$ systems}\label{sec:app}

\begin{figure}
	\includegraphics[width=1.0\linewidth]{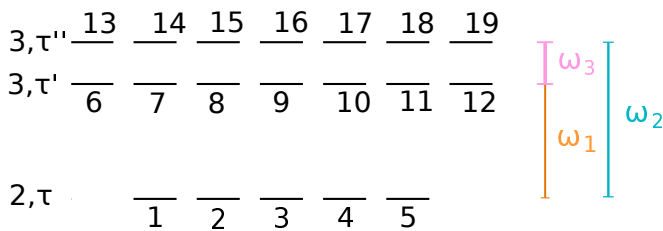} 
\caption{Level scheme for a rotational subsystem consisting of the levels $J=2,\tau$, $J=3, \tau'$, and $J=3, \tau''$. The frequencies of the control fields are $\omega_1$, $\omega_2$, and $\omega_3$.}
	\label{schemeJ233}
\end{figure}
As a further example for the controllability of asymmetric top rotors, we prove enantio-selective controllability for subsystems consisting of rotational levels with $J=2$, $J=3$, $J=3$, as shown in Fig.~\ref{schemeJ233}. 
The relevant Hilbert space is $\mathcal{H}^{(+)}\oplus \mathcal{H}^{(-)}$, where 
\begin{eqnarray*}
\mathcal{H}^{(\pm)}&=&\mathrm{span}\{|2\tau M\rangle^{(\pm)}\mid M=-2,\dots, 2\}\\ && \oplus \mathrm{span}\{|3\tau' M\rangle^{(\pm)},|3\tau''M\rangle^{(\pm)}  \mid M=-3,\dots,3\,\},  
\end{eqnarray*}
with $\tau \in \{-2,\dots,2\}$, and $\tau',\tau'' \in \{-3,\dots,3\}$ is the subspace of rotational eigenstates for each of the enantiomers, and $\mathcal{H}^{(+)}\oplus\mathcal{H}^{(-)}\cong \mathbb{C}^{19}\oplus \mathbb{C}^{19}$. 
The rotational Hamiltonian of a single enantiomer is given by Eq.~\eqref{Ham_int_tot} with 
\[
\begin{split}
{\bf H}_{0}&=\mathrm{diag}(E_{2,\tau},E_{2,\tau},E_{2,\tau},E_{2,\tau},E_{2,\tau},\\ &
E_{3,\tau'},E_{3,\tau'},E_{3,\tau'},E_{3,\tau'},E_{3,\tau'},E_{3,\tau'},E_{3,\tau'},\\ &
E_{3,\tau''},E_{3,\tau''},E_{3,\tau''},E_{3,\tau''},E_{3,\tau''},E_{3,\tau''},E_{3,\tau''})\,,
\end{split}
\]
where $E_{2,\tau},E_{3,\tau'},E_{3,\tau''}$ are the rotational energy levels of the asymmetric top. 
In the following, we prove that the set of five control fields that results in the interaction Hamiltonians
$$\mathcal{X}=\{\mathrm{i}{\bf H}_{\omega_1,x}^{chiral},\mathrm{i}{\bf H}_{\omega_1,y}^{chiral},\mathrm{i}{\bf H}_{\omega_2,y}^{chiral},\mathrm{i}{\bf H}_{\omega_2,z}^{chiral},\mathrm{i}{\bf H}_{\omega_3,x}^{chiral}\}$$
with $\omega_1=|E_{3,\tau'}-E_{2,\tau}|$, $\omega_2=|E_{3,\tau''}-E_{3,\tau'}|$, and $\omega_3=|E_{3,\tau''}-E_{2,\tau}|$,
yields complete enantio-selective controllability.
We assume here, without loss of generality, that the transition with $\omega_1$
($\omega_2$, respectively $\omega_3$) couples to $\mu_c$ ($\mu_a$, respectively $\mu_b$) and that  $\mu_b^{(-)}=-\mu_b^{(+)}$. The frequencies are also indicated in Fig.~\ref{schemeJ233}.  

We first prove that the set of interaction operators for four control fields, 
$$\mathcal{X}_1=\{\mathrm{i}{\bf H}_{\omega_1,x},\mathrm{i}{\bf H}_{\omega_1,y},\mathrm{i}{\bf H}_{\omega_2,y},\mathrm{i}{\bf H}_{\omega_2,z}\}\,,$$
together with the rotational Hamiltonian, ${\bf H}_{0}$, gives controllability for a single enantiomer. To this end, we have to show that 
\begin{equation}\label{su(19)}
\mathrm{Lie}\{ \{ \mathrm{i}{\bf H}_{0}\}\cup\mathcal{X}_1 \}\}=\mathfrak{su}(19),
\end{equation}
since the Hilbert space for each enantiomer $\mathcal{H}^{(\pm)}$ coincides with $\mathbb{C}^{19}$. 
We express the interaction Hamiltonians in terms of the generalized Pauli matrices (\ref{basis}), 
\begin{widetext}
  \begin{eqnarray*}
	\mathrm{i}{\bf H}_{\omega_1,x}&\propto&\mu_c\Big(\sqrt{15}({\bf G}_{1,6}+{\bf G}_{5,12})+\sqrt{10}({\bf G}_{2,7}+{\bf G}_{4,11})+\sqrt{6}({\bf G}_{3,8}+{\bf G}_{3,10})+\sqrt{3}({\bf G}_{2,9}+{\bf G}_{4,9})+({\bf G}_{1,8}+{\bf G}_{5,10})\Big),\\
	\mathrm{i}{\bf H}_{\omega_1,y}&\propto&\mu_c\Big(\sqrt{15}(-{\bf F}_{1,6}+{\bf F}_{5,12})+\sqrt{10}(-{\bf F}_{2,7}+{\bf F}_{4,11})+\sqrt{6}(-{\bf F}_{3,8}+{\bf F}_{3,10})+\sqrt{3}({\bf F}_{2,9}-{\bf F}_{4,9})+({\bf F}_{1,8}-{\bf F}_{5,10})\Big),\\
    \mathrm{i}{\bf H}_{\omega_2,y}&\propto&\mu_a\Big(\sqrt{15}(-{\bf G}_{1,13}+{\bf G}_{5,19})+\sqrt{10}(-{\bf G}_{2,14}+{\bf G}_{4,18})+\sqrt{6}(-{\bf G}_{3,5}+{\bf G}_{3,17})+\sqrt{3}(-{\bf G}_{4,16}
                                            +{\bf G}_{2,16})\\ &&\quad\qquad+(-{\bf G}_{5,17}+{\bf G}_{1,15})\Big),\\ 
	\mathrm{i}{\bf H}_{\omega_2,z}&\propto&\mu_a\Big(\sqrt{8}({\bf G}_{2,15}+{\bf G}_{4,17})+\sqrt{5}({\bf G}_{1,14}+{\bf G}_{5,18})+{\bf G}_{3,16}\Big)\,.
        \end{eqnarray*}
Using the relations \eqref{eq:commutators}, 
we compute
\begin{eqnarray*}
	[\mathrm{i}{\bf H}_{0},\mathrm{i}{\bf H}_{\omega_1,y}] &\propto& \sqrt{15}(-{\bf G}_{1,6}+{\bf G}_{5,12})+\sqrt{10}(-{\bf G}_{2,7}+{\bf G}_{4,11})+\sqrt{6}(-{\bf G}_{3,8}+{\bf G}_{3,10})\\&&+\sqrt{3}({\bf G}_{2,9}-{\bf G}_{4,9})+({\bf G}_{1,8}-{\bf G}_{5,10})\\&=:&J(\mathrm{i}{\bf H}_{\omega_1,y})\,.  
\end{eqnarray*}
Next, we consider the Hamiltonian
\begin{eqnarray}
\mathrm{i} {\bf H}_{\omega_1,\sigma_+} &:=& J(\mathrm{i}{\bf H}_{\omega_1,y})+\mathrm{i}{\bf H }_{\omega_1,x} 
\propto
\sqrt{15}{\bf G}_{5,12} +\sqrt{10}{\bf G}_{4,11}
+\sqrt{6}{\bf G}_{3,10}+\sqrt{3}{\bf G}_{2,9}+{\bf G}_{1,8}\,,
\end{eqnarray}
which corresponds to the interaction with a right circularly polarized field with frequency $\omega_1$.
Defining $J(\mathrm{i}{\bf H}_{\omega_1,\sigma_+})=[\mathrm{i}{\bf H}_{0} ,\mathrm{i}{\bf H}_{\omega_1,\sigma_+}]/\omega_1$, 
we find 
\[
\mathrm{ad}^{2s}_{J({\rm i}{\bf H}_{\omega_1,\sigma_+})} {\rm i}{\bf H}_{\omega_1,\sigma_+}\propto
\sqrt{15}^{2s+1}{\bf G}_{5,12}+\sqrt{10}^{2s+1}{\bf G}_{4,11}+\sqrt{6}^{2s+1}{\bf G}_{3,10}+\sqrt{3}^{2s+1}{\bf G}_{2,9}+{\bf G}_{1,8}\,.
\]
      \end{widetext}
We can thus write 
	$$\begin{pmatrix}
	\mathrm{ad}^{0}_{J({\rm i}{\bf H}_{\omega_1,\sigma_+})} {\rm i}{\bf H}_{\omega_1,\sigma_+} \\
	\mathrm{ad}^{2}_{J({\rm i}{\bf H}_{\omega_1,\sigma_+})} {\rm i}{\bf H}_{\omega_1,\sigma_+} \\
	\mathrm{ad}^{4}_{J({\rm i}{\bf H}_{\omega_1,\sigma_+})} {\rm i}{\bf H}_{\omega_1,\sigma_+}\\
	\mathrm{ad}^{6}_{J({\rm i}{\bf H}_{\omega_1,\sigma_+})} {\rm i}{\bf H}_{\omega_1,\sigma_+}\\
	\mathrm{ad}^{8}_{J({\rm i}{\bf H}_{\omega_1,\sigma_+})} {\rm i}{\bf H}_{\omega_1,\sigma_+}
	\end{pmatrix}=V \begin{pmatrix}
	{\bf G}_{5,12}\\
	{\bf G}_{4,11}\\
	{\bf G}_{3,10}\\
	{\bf G}_{2,9}\\
	{\bf G}_{1,8} 
	\end{pmatrix}
	$$
with
  $$
	\qquad V=\begin{pmatrix}
	\sqrt{15} & \sqrt{10} &\sqrt{6} &\sqrt{3}& 1\\
	\sqrt{15}^{3} & \sqrt{10}^{3} &\sqrt{6}^{3} &\sqrt{3}^{3}& 1\\
	\sqrt{15}^{5} & \sqrt{10}^{5} &\sqrt{6}^{5} &\sqrt{3}^{5}& 1\\
	\sqrt{15}^{7} & \sqrt{10}^{7} &\sqrt{6}^{7} &\sqrt{3}^{7}& 1\\
	\sqrt{15}^{9} & \sqrt{10}^{9} &\sqrt{6}^{9} &\sqrt{3}^{9}& 1
	\end{pmatrix}. $$
Since $V$ is a Vandermonde matrix,  its determinant is given by the product of the sum and the difference of the coefficients of the first row.
Since those coefficients are all different, $V$ is invertible, and thus
\begin{equation}\label{elements1}
{\bf G}_{5,12}, {\bf G}_{4,11}, {\bf G}_{3,10}, {\bf G}_{2,9}, {\bf G}_{1,8} \in \mathrm{Lie}\{\mathrm{i}{\bf H}_{0} ,\mathrm{i}{\bf H}_{\omega_1,\sigma_+}\}.
\end{equation}
With a completely analogous argument, we obtain that 
\begin{equation}\label{elements2}
\begin{split}
&(-{\bf G}_{1,13}+ {\bf G}_{5,19}),(-{\bf G}_{2,14}+ {\bf G}_{4,18}),
(-{\bf G}_{3,15}+ {\bf G}_{3,17}), \\ & (-{\bf G}_{4,16}+ {\bf G}_{2,16}),
(-{\bf G}_{5,17}+ {\bf G}_{1,15}) \in \mathrm{Lie}\{\mathrm{i}{\bf H}_{0} ,\mathrm{i}{\bf H}_{\omega_2,y}\}.
\end{split}
\end{equation}
Using the basis elements \eqref{elements1}, we can break the sums into the  elements \eqref{elements2}, e.g., 
$$[[-{\bf G}_{1,13}+ {\bf G}_{5,19},{\bf G}_{1,8}],{\bf G}_{1,8}]\propto {\bf G}_{1,13}. $$
Moreover, commutators between the elements \eqref{elements1} and $\mathrm{i}{\bf H}_{\omega_2,z}$ generate 
$$[[\mathrm{i}{\bf H}_{\omega_2,z},{\bf G}_{1,8}],{\bf G}_{1,8}]\propto {\bf G}_{1,14},$$
and all other ${\bf G}_{i,j}$ occurring in $\mathrm{i} {\bf H}_{\omega_2,z}$. 
Finally, we repeat the previous calculations with the interaction operator
\begin{eqnarray*}
{\rm i}{\bf H}_{\omega_1,\sigma_-}&:=&J(\mathrm{i}{\bf H}_{\omega_1,y})-\mathrm{i}{\bf H }_{\omega_1,x}\\&\propto&
\sqrt{15}{\bf G}_{1,6}+\sqrt{10}{\bf G}_{2,7}+\sqrt{6}{\bf G}_{3,8}+\sqrt{3}{\bf G}_{4,9}\\&&+{\bf G}_{5,10} \,,	
\end{eqnarray*}
which corresponds to the interaction with a  left circular polarized field, and obtain 
$${\bf G}_{1,6},{\bf G}_{2,7},{\bf G}_{3,8},{\bf G}_{4,9},{\bf G}_{5,10}\in   \mathrm{Lie}\{\mathrm{i}{\bf H}_{0} ,{\rm i}{\bf H}_{\omega_1,\sigma_-}\}.$$
With the relations~\eqref{eq:commutators}, 
we find all remaining basis elements and thus prove \eqref{su(19)}.

As a second step, we add a fifth field with interaction Hamiltonian $\mathrm{i}{\bf H}_{\omega_3,x}$ in order to obtain enantio-selective controllability in $\mathcal{H}^{(+)}\oplus \mathcal{H}^{(-)}$. The interaction Hamiltonian for the composite system becomes 
\begin{eqnarray}
\mathrm{i}{\bf H}_{\omega_3,x}^{chiral}= \left( 
\begin{array}{cc}
\mathrm{i}{\bf H}_{\omega_3,x} & 0  \\
0 & -\mathrm{i}{\bf H}_{\omega_3,x}  \end{array} \right) \,,
\end{eqnarray}
where
\begin{widetext}
  \begin{equation}\label{5th}
    \begin{split}
      \mathrm{i}{\bf H}_{\omega_3,x}\propto\mu_b\Big(&\sqrt{3}({\bf F}_{6,14}+{\bf F}_{7,13}+{\bf F}_{11,19}+{\bf F}_{12,18})+\sqrt{5}({\bf F}_{7,15}+{\bf F}_{8,14}+{\bf F}_{10,18}+{\bf F}_{11,17})+\\ &\sqrt{6}({\bf F}_{8,16}+{\bf F}_{9,15}+{\bf F}_{9,17}+{\bf F}_{10,16}) \Big).
    \end{split}
  \end{equation}
\end{widetext}
The extension of the action of the previous four interaction Hamiltonians on both enantiomers is given by a direct sum, 
$$\mathrm{i}{\bf H}_{\omega,a}^{chiral}= \begin{pmatrix}
\mathrm{i}{\bf H}_{\omega,a} & 0\\
0 & \mathrm{i}{\bf H}_{\omega,a}
\end{pmatrix} $$
for $\omega=\omega_1$ and $\omega=\omega_2$. Therefore, the set of operators $\mathcal{X}_1$ and the rotational Hamiltonian ${\bf H}_0$ generate the Lie algebra 
$$\Big\{\begin{pmatrix}
A & 0\\
0 & A
\end{pmatrix} \mid A\in \mathfrak{su}(19) \Big\}\cong \mathfrak{su}(19),$$
as operators acting on the space of states $\mathcal{H}^{(+)}\oplus \mathcal{H}^{(-)}=\mathbb{C}^{19}\oplus\mathbb{C}^{19}$.
Now, regarding the coefficients in Eq.~\eqref{5th}, we can separate the sum with respect to $\sqrt{3}$ (and also with respect to $\sqrt{5}$ and $\sqrt{6}$) using a usual Vandermonde argument, obtaining
\begin{widetext}
	$$\begin{pmatrix}
	{\bf G}_{6,14}+{\bf G}_{12,18}+{\bf G}_{7,13}+{\bf G}_{11,19}&0\\
	0&-({\bf G}_{6,14}+{\bf G}_{12,18}+{\bf G}_{7,13}+{\bf G}_{11,19})
	\end{pmatrix} \in \mathrm{Lie}\{\mathrm{i}{\bf H}_{\omega_3,x}^{chiral},\mathrm{i}{\bf H}_{0}^{chiral}\}.$$ 
	Since the element 
	$\begin{pmatrix}
	{\bf G}_{6,13}&0\\
	0&{\bf G}_{6,13}
	\end{pmatrix} $ belongs to the generated Lie algebra because of the first step, we can compute the following commutators inside the generated Lie algebra,
	\[
	\begin{split}
	&\mathrm{ad}^2_{\begin{pmatrix}
		{\bf G}_{6,13}&0\\
		0&{\bf G}_{6,13}
		\end{pmatrix}}\left(\begin{pmatrix}
	{\bf G}_{6,14}+{\bf G}_{12,18}+{\bf G}_{7,13}+{\bf G}_{11,19}&0\\
	0&-({\bf G}_{6,14}+{\bf G}_{12,18}+{\bf G}_{7,13}+{\bf G}_{11,19})
	\end{pmatrix} \right)
	\\ &\propto \begin{pmatrix}
	({\bf G}_{6,14}+{\bf G}_{7,13}) & 0\\
	0& -({\bf G}_{6,14}+{\bf G}_{7,13})
	\end{pmatrix}\,.
	\end{split}
	\]
\end{widetext}
Since $\begin{pmatrix}
{\bf G}_{6,14}+{\bf G}_{7,13} & 0 \\
0 & {\bf G}_{6,14}+{\bf G}_{7,13}
\end{pmatrix}$ is in the generated Lie algebra (because of the first step), 
$$ \begin{pmatrix}
{\bf G}_{6,14}+{\bf G}_{7,13} & 0 \\
0 & 0
\end{pmatrix}, \quad \begin{pmatrix}
0 & 0 \\
0 & {\bf G}_{6,14}+{\bf G}_{7,13}
\end{pmatrix}$$
are also in the generated Lie algebra. These two operators  break the parity between enantiomers, since they act on the first and on the second enantiomer only. We are thus left to prove that we can separate the elements ${\bf G}_{6,14}$ and ${\bf G}_{7,13}$. To this end, it suffices to consider the double commutator inside the generated Lie algebra
\[
\begin{split}
&\Bigg[\Bigg[\begin{pmatrix}
{\bf G}_{6,14}+{\bf G}_{7,13} & 0 \\
0 & 0
\end{pmatrix}, \begin{pmatrix}
{\bf F}_{6,14} & 0 \\
0 & {\bf F}_{6,14}
\end{pmatrix}\Bigg],\begin{pmatrix}
{\bf F}_{6,14} & 0 \\
0 & {\bf F}_{6,14}
\end{pmatrix} \Bigg]\\ &\propto \begin{pmatrix}
{\bf G}_{6,14} & 0 \\
0 & 0
\end{pmatrix}.
\end{split}
\]
Using the relations~\eqref{eq:commutators}, we furthermore see 
that all the other basis elements belong to the generated Lie algebra, which  proves that 
\begin{widetext}
	$$\mathrm{Lie}\{ \{ \mathrm{i}{\bf H}_{0}^{chiral}\}\cup\mathcal{X}\}=\mathrm{span}\Big\{ \begin{pmatrix}
	A & 0\\
	0 & 0
	\end{pmatrix}, \begin{pmatrix}
	0 & 0\\
	0 & A
	\end{pmatrix} \mid A\in \mathfrak{su}(19)\Big\}\cong\mathfrak{su}(19)\oplus \mathfrak{su}(19)\,,  $$
\end{widetext}
implying that the $J=2/3/3$-system is enantio-selective controllable.

%


\end{document}